\documentclass[sigconf]{acmart} %

\usepackage{xspace} %
\usepackage[shortlabels]{enumitem} %
\usepackage{stfloats} %

\AtBeginDocument{%
  }

\copyrightyear{2026}
\acmYear{2026}
\setcopyright{cc}
\setcctype{by-nc-nd}
\acmConference[CHI '26]{Proceedings of the 2026 CHI Conference on Human Factors in Computing Systems}{April 13--17, 2026}{Barcelona, Spain}
\acmBooktitle{Proceedings of the 2026 CHI Conference on Human Factors in Computing Systems (CHI '26), April 13--17, 2026, Barcelona, Spain}
\acmDOI{10.1145/3772318.3791337}
\acmISBN{979-8-4007-2278-3/2026/04}

\begin{document}

\title{Vidmento: Creating Video Stories Through Context-Aware Expansion With Generative Video}

\author{Catherine Yeh}
\orcid{0009-0007-0429-4770}
\authornote{Work done at Adobe.}
\affiliation{%
  \institution{Harvard University}
  \city{Boston}
  \state{MA}
  \country{USA}
}
\email{catherineyeh@g.harvard.edu}

\author{Anh Truong}
\orcid{0009-0005-5409-7287}
\affiliation{%
  \institution{Adobe Research}
  \city{New York}
  \state{NY}
  \country{USA}}
\email{truong@adobe.com}

\author{Mira Dontcheva}
\orcid{0009-0006-5394-2706}
\affiliation{%
  \institution{Adobe Research}
  \city{Seattle}
  \state{WA}
  \country{USA}}
\email{mirad@adobe.com}

\author{Bryan Wang}
\orcid{0000-0001-9016-038X}
\affiliation{%
  \institution{Adobe Research}
  \city{Seattle}
  \state{WA}
  \country{USA}}
\email{bryanw@adobe.com}

\renewcommand{\shortauthors}{Yeh et al.}

\definecolor{RoyalBlue}{HTML}{0071BC}
\newcommand{\todo}[1]{{\textcolor{orange}{[TODO: #1]}\normalfont}}
\newcommand{\cy}[1]{{\textcolor{magenta}{[CY: #1]}\normalfont}}
\newcommand{\bw}[1]{{\textcolor{blue}{[BW: #1]}\normalfont}}
\newcommand{\mira}[1]{{\textcolor{purple}{[MD: #1]}\normalfont}}

\newcommand{\ie}{{i.e.,}\xspace}
\newcommand{\eg}{{e.g.,}\xspace}
\newcommand{\ea}{{et~al\xperiod}\xspace}
\newcommand{\aka}{{a.k.a.}\xspace}
\newcommand{\etc}{{etc\xperiod}\xspace}
\newcommand{\etal}{{et al\xperiod}\xspace}

\newcommand{\system}{\textsc{Vidmento}\xspace}
\newcommand{\systemx}{\textsc{Vidmento}}

\newcommand{\location}{Adobe\xspace}

\newcommand{\numbooks}{9\xspace}

\newcommand{\explore}{\textbf{D1: \textit{Explore}}}
\newcommand{\expand}{\textbf{D2: \textit{Expand}}}
\newcommand{\blend}{\textbf{D3: \textit{Blend}}}
\newcommand{\control}{\textbf{D4: \textit{Control}}}

\renewcommand{\sectionautorefname}{Section}
\renewcommand{\subsectionautorefname}{Section}
\renewcommand{\subsubsectionautorefname}{Section}

\definecolor{AIColor}{HTML}{CC8D6C}
\definecolor{RealColor}{HTML}{939BC0}

\definecolor{GreenColor}{HTML}{98AFB5}
\definecolor{PinkColor}{HTML}{CD85BA}
\definecolor{OrangeColor}{HTML}{CC8D6C}

\definecolor{Gray}{HTML}{777777}

\newcommand{\myquote}[1]{\emph{``#1''}}
\newcommand{\boldpar}[1]{\paragraph{\textbf{#1}}}

\newcommand{\gray}[1]{\textcolor{Gray}{#1}}

\newcommand{\llm}{\includegraphics[width=0.12cm]{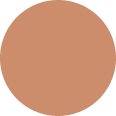}\xspace}
\newcommand{\image}{\includegraphics[width=0.12cm]{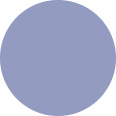}\xspace}
\newcommand{\video}{\includegraphics[width=0.12cm]{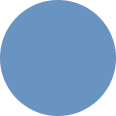}\xspace}
\newcommand{\narrate}{\includegraphics[width=0.12cm]{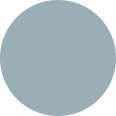}\xspace}
\newcommand{\music}{\includegraphics[width=0.12cm]{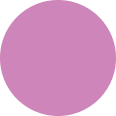}\xspace}

\newcommand{\figureTeaser}{
\begin{teaserfigure}
  \includegraphics[width=\textwidth]{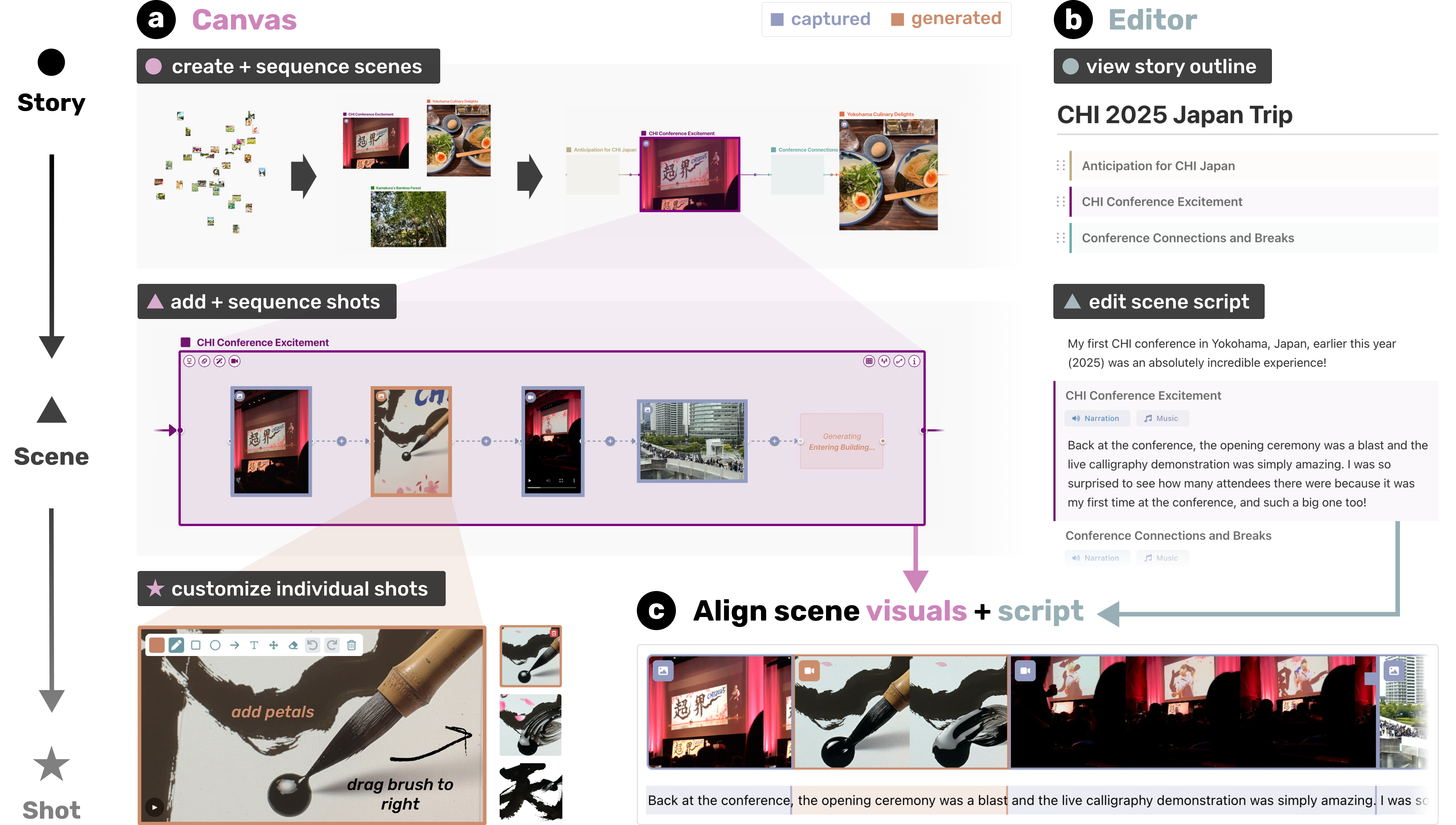}
  \caption{\textbf{We introduce \system, a video authoring tool} that helps creators craft compelling video stories (\eg a vlog about CHI 2025) 
  through augmenting their initial footage and ideas with contextual video generations.
  \system allows creators to explore and edit their video at different narrative levels (story, scene, shot) through two linked views: the \textbf{(a)} \textit{canvas} and \textbf{(b)} \textit{editor}.
  We also designed a \textbf{(c)} \textit{timeline} for directly aligning visuals to the script.
  Captured (existing) shots are denoted by a \textcolor{RealColor}{purple} outline, while new generated media are outlined in \textcolor{AIColor}{orange}.
  }
  \Description{Figure showing the Vidmento interface with three panels: (a) a canvas for creating and sequencing story scenes, (b) an editor for writing and editing scene scripts, and (c) a timeline for aligning visuals with the script. Visuals are organized hierarchically: stories at the top, then scenes, then individual shots. Captured media is outlined in purple; generated media is outlined in orange.}
  \label{fig:teaser}
\end{teaserfigure}
}

\newcommand{\figureScaffolded}{
\begin{figure*}
  \includegraphics[width=\textwidth]{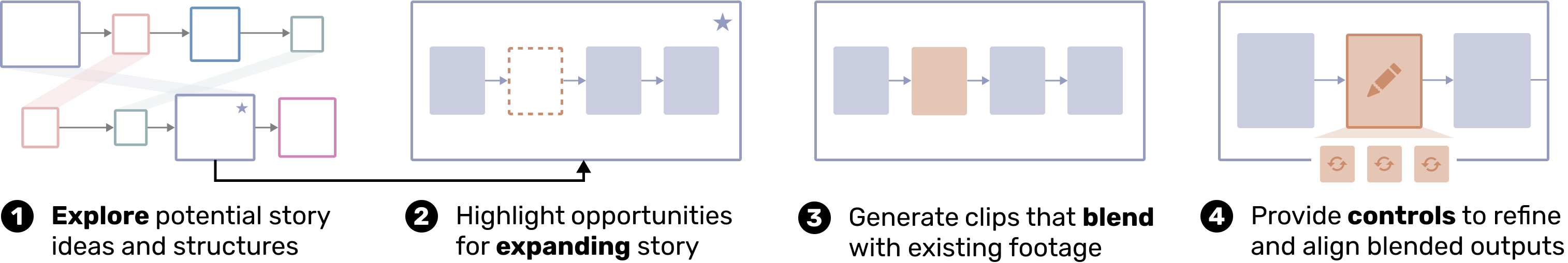}
  \caption{\textbf{Overview of generative expansion}, a design framework for contextually expanding existing footage with the affordances of generative video. 
  This approach is designed to provide structured support for video storytelling, while offering flexibility for creative exploration and iteration. 
  }
  \Description{Diagram illustrating Vidmento’s “generative expansion” workflow for video storytelling. Four steps are shown: (1) explore potential story ideas and structures, (2) highlight opportunities for expanding the story, (3) generate clips that blend with existing footage, and (4) provide controls to refine and align blended outputs.}
  \label{fig:scaffolded}
\end{figure*}
}

\newcommand{\figureInterface}{
\begin{figure*}[t]
    \centering
    \includegraphics[width=0.91\linewidth]{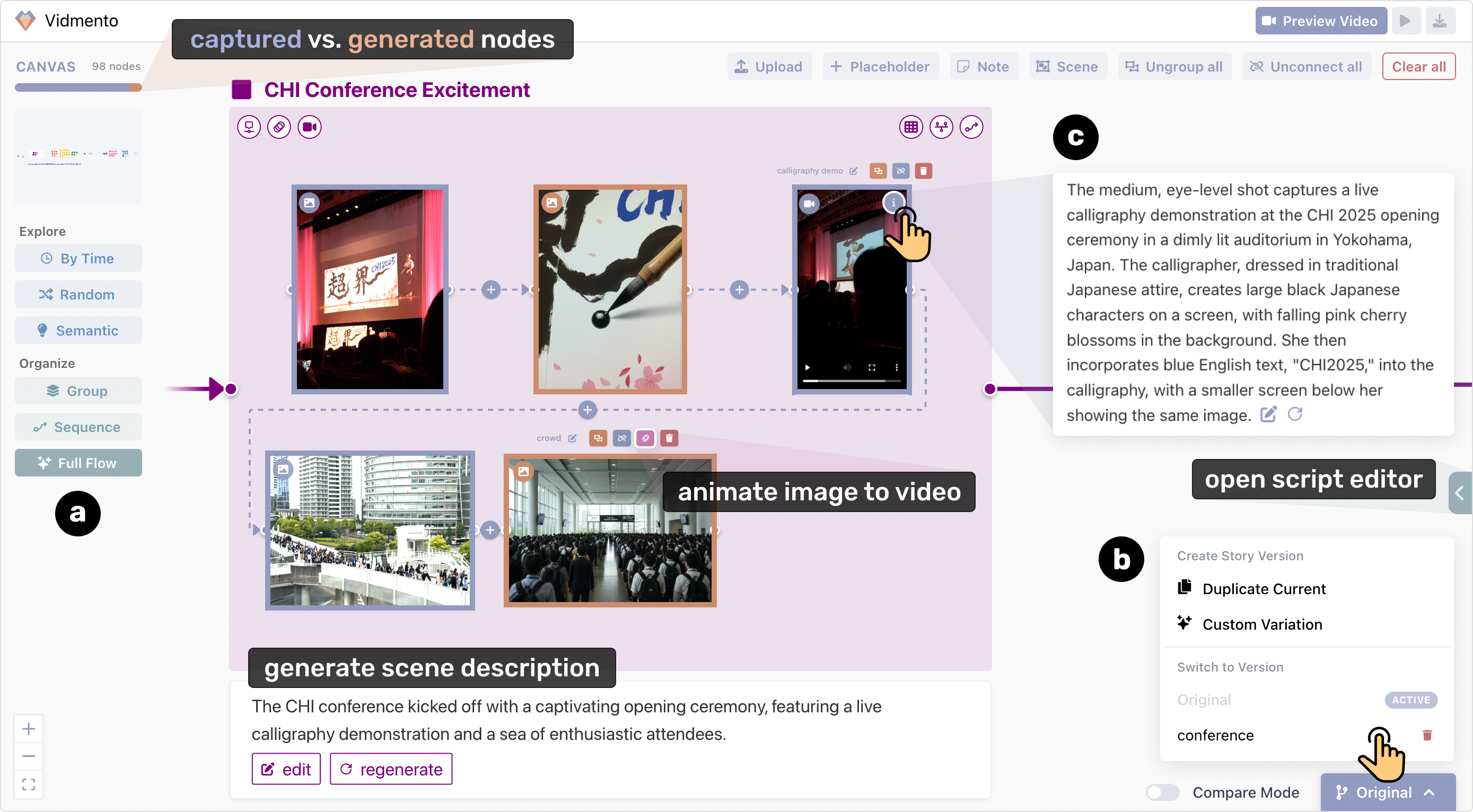}
    \caption{\textbf{\system's canvas view}. \textbf{(a)} To the left, there are different options for exploring and organizing visual materials. \textbf{(b)} Users can create and compare different story versions by duplicating the current version or prompting for a custom variation. \textbf{(c)} Clicking on a file node (\textcolor{RealColor}{captured} or \textcolor{AIColor}{generated}) reveals its description, which can be edited or regenerated. 
    }
    \Description{Screenshot of Vidmento’s canvas view. (a) Left sidebar offers options to explore and organize media. (b) Users can duplicate or create variations of story versions. (c) Clicking on a node reveals descriptions that can be edited or regenerated. Example shows a scene with captured and generated clips sequenced together. 
    }
    \label{fig:interface}
\end{figure*}
}

\newcommand{\figureEditor}{
\begin{figure*}[t]
    \centering
    \includegraphics[width=0.91\linewidth]{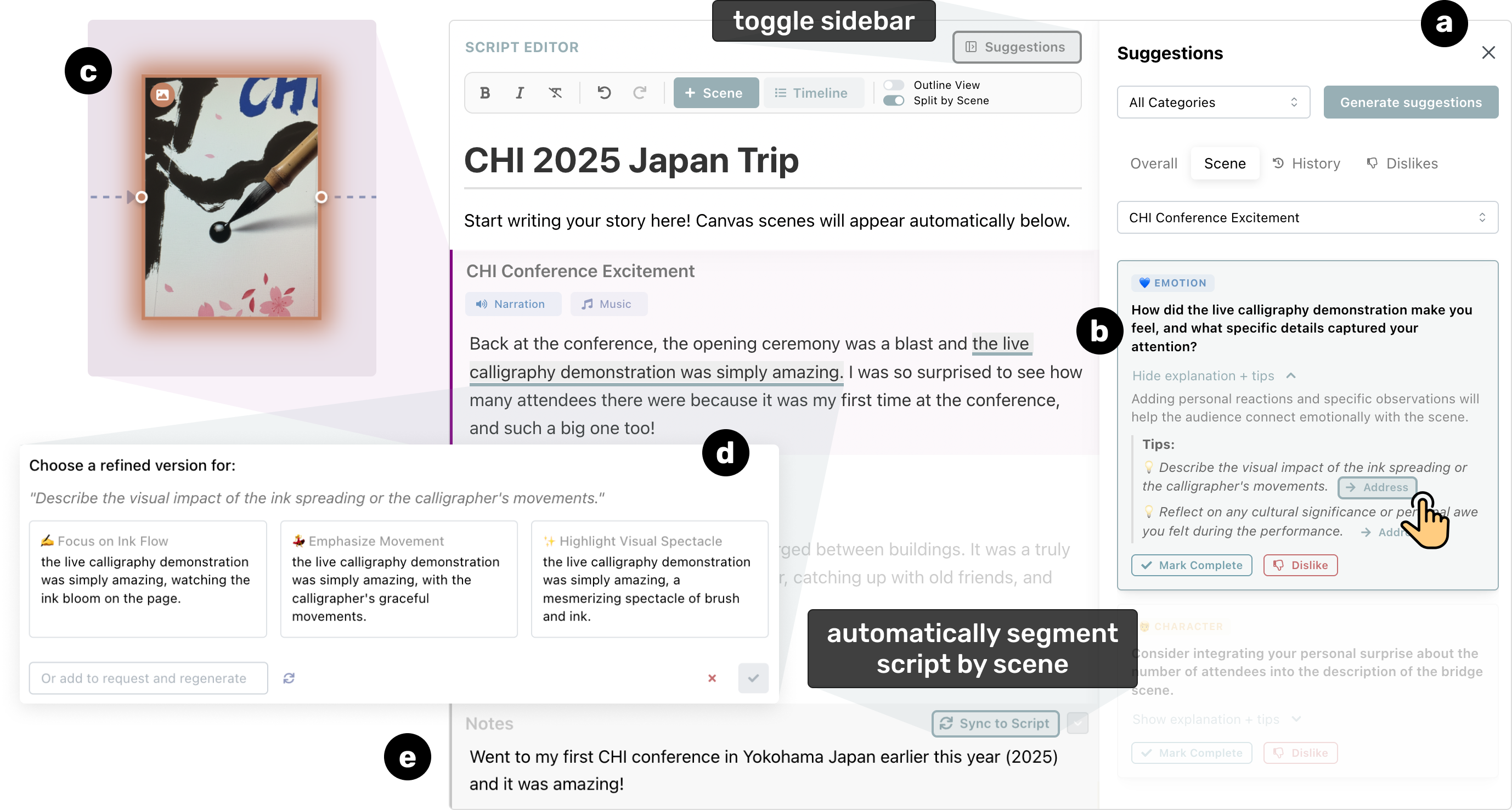}
    \caption{\textbf{\system's script editor}. \textbf{(a)} Suggestions for script expansion appear in the sidebar. \textbf{(b)} Each includes an explanation and tips for addressing it. \textbf{(c)} Clicking ``address'' highlights relevant visuals in the canvas and 
    \textbf{(d)} suggests modifications via the inline refinement menu. 
    \textbf{(e)} Users can add ideas or story context in the editor \textit{notes} to steer system generations. 
    }
    \Description{Screenshot of Vidmento’s script editor. (a) Suggestions sidebar provides prompts to expand writing. (b) Suggestions include explanations and tips. (c) Clicking “address” links script to visuals. (d) Inline refinement offers phrasing variations. (e) Notes section allows users to capture additional ideas. The editor automatically segments text by scene.
    }
    \label{fig:editor}
\end{figure*}
}

\newcommand{\figureGroup}{
\begin{figure}[t]
    \centering
    \includegraphics[width=\linewidth]{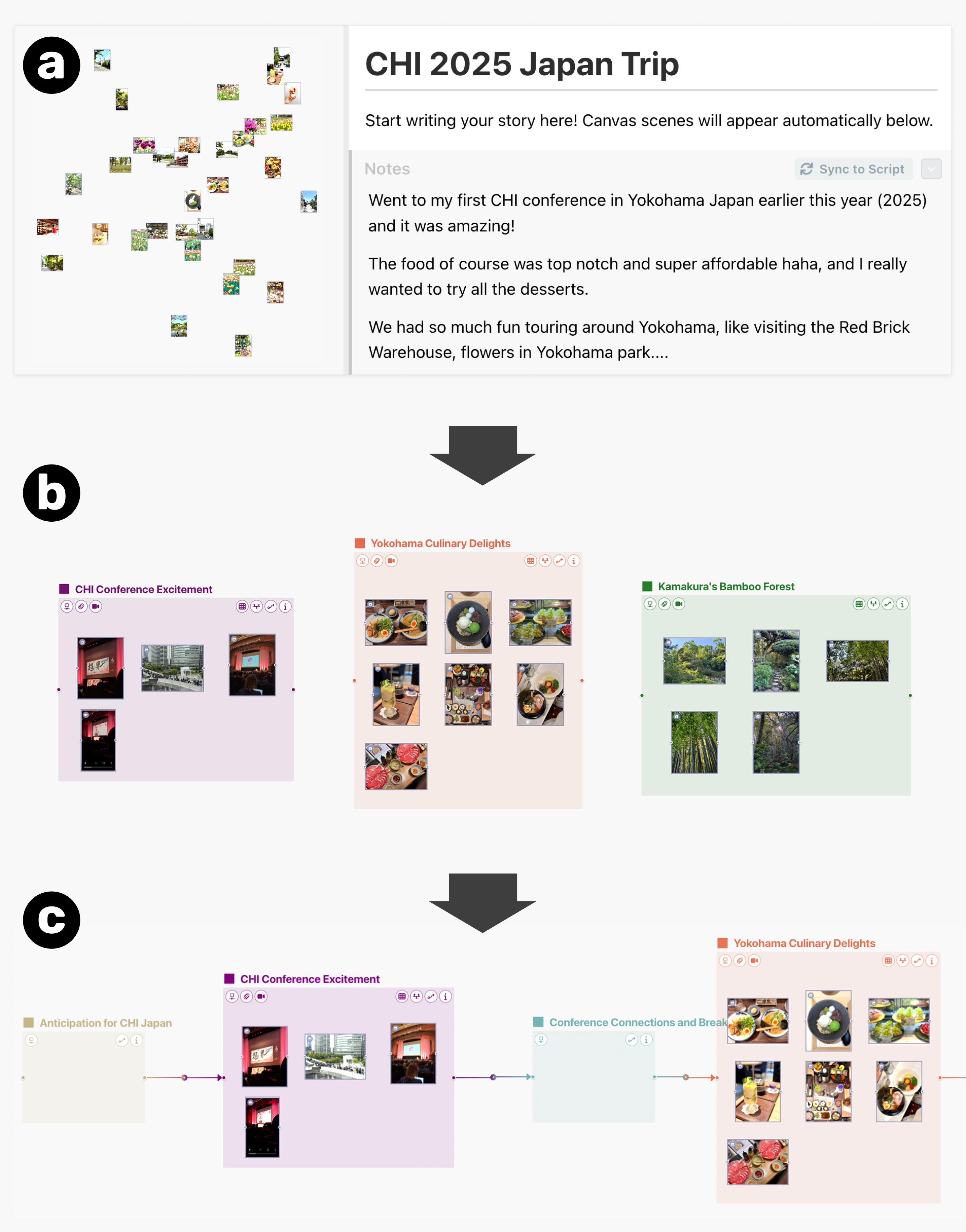}
    \caption{\textbf{Shaping a video story} with \system from \textbf{(a)} an initial set of materials by \textbf{(b)} grouping shots into scenes and 
    \textbf{(c)} connecting scenes into a sequence.
    }
    \Description{Illustration of shaping a video story. (a) Initial scattered set of media files. (b) Grouping files into scenes with titles and descriptions. (c) Sequencing the scenes into a connected story outline. Example shows scenes about CHI conference events, food, and sightseeing.}
    \label{fig:group}
\end{figure}
}

\newcommand{\figureCompare}{
\begin{figure*}[ht]
    \centering
    \includegraphics[width=\linewidth]{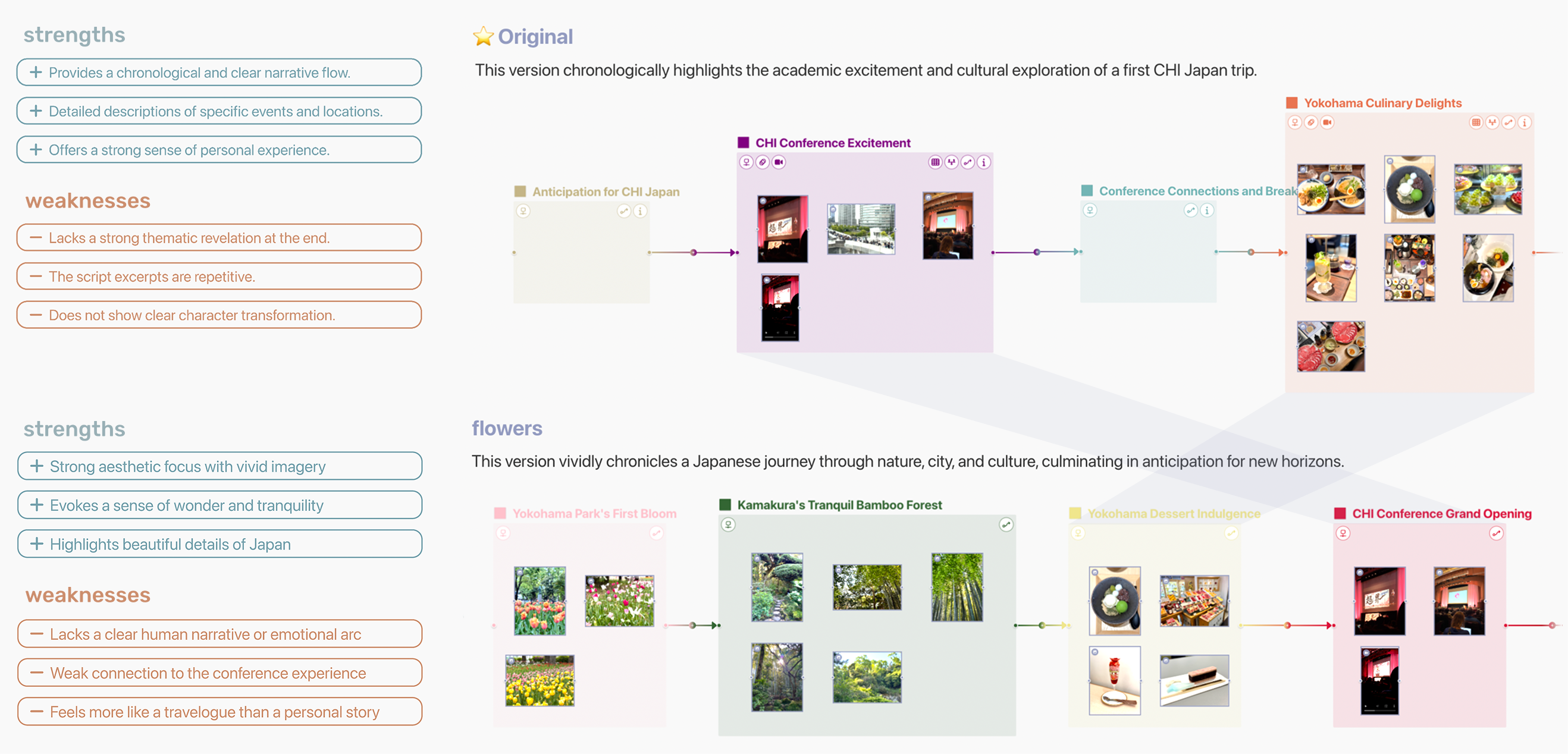}
    \caption{\textbf{Comparing story versions in compare mode.} For each variation, a brief summary is provided, along with a list of \textcolor{GreenColor}{strengths} and \textcolor{OrangeColor}{weaknesses}. To facilitate comparison, scenes at the same position across story versions are aligned horizontally.}
    \Description{Screenshot of compare mode for viewing different story versions. Top version (“Original”) presents chronological highlights; bottom version (“Flowers”) emphasizes aesthetic imagery. Strengths and weaknesses of each version are listed, such as narrative clarity versus weaker emotional arc.}
    \label{fig:compare}
\end{figure*}
}

\newcommand{\figureNewScene}{
\begin{figure}[ht]
    \centering
    \includegraphics[width=0.6\linewidth]{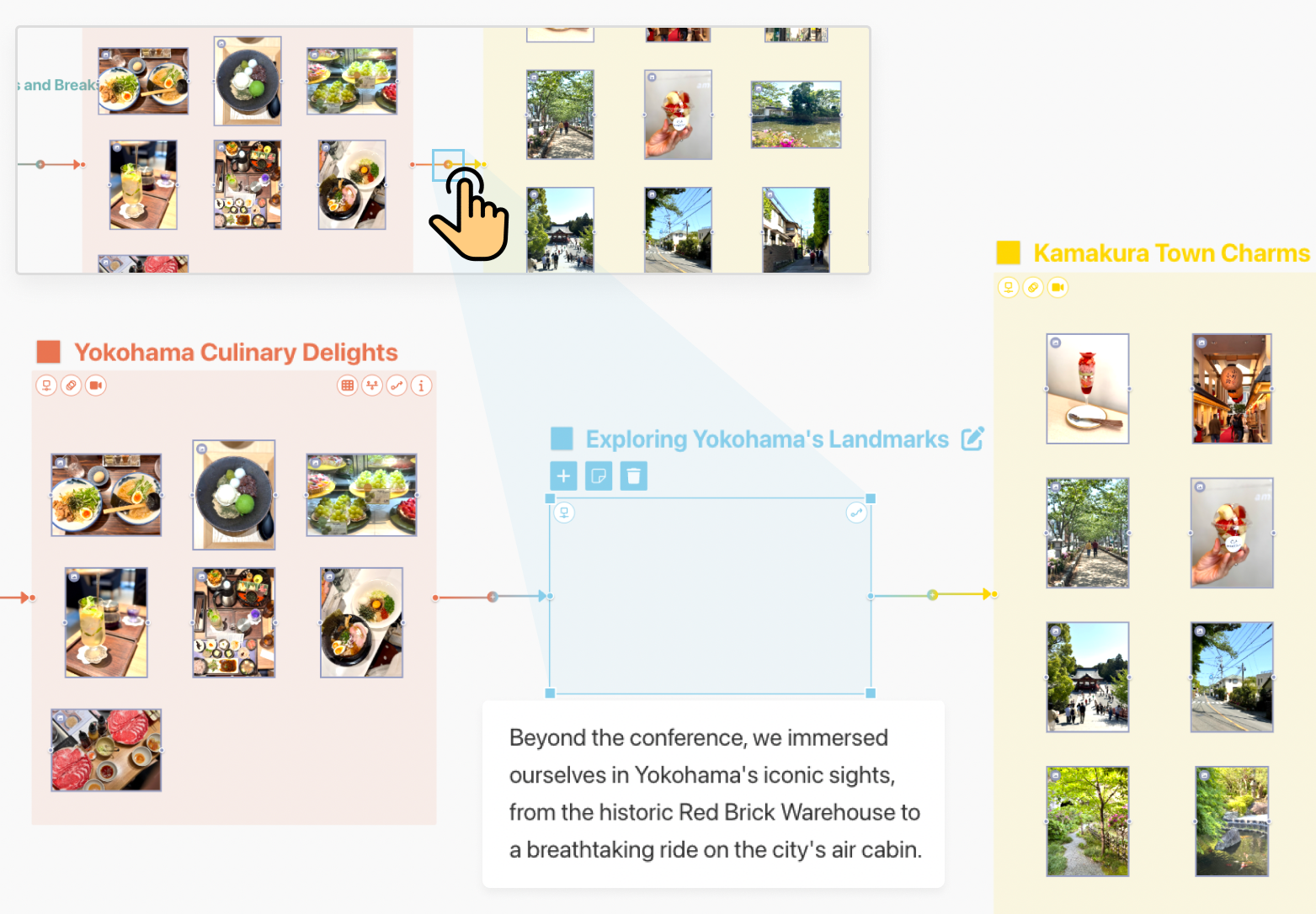}
    \caption{\textbf{Adding a contextual scene} by pressing the plus button on an edge between two existing scenes.}
    \Description{Screenshot showing how Vidmento helps create contextual transition scenes. By pressing a plus button between two scenes, the system suggests a connecting scene with description. Example shows transitions between “Yokohama Culinary Delights” and “Kamakura Town Charms.”}
    \label{fig:new_scene}
\end{figure}
}
\newcommand{\figureVisualSequence}{
\begin{figure*}[t]
    \centering
    \includegraphics[width=\linewidth]{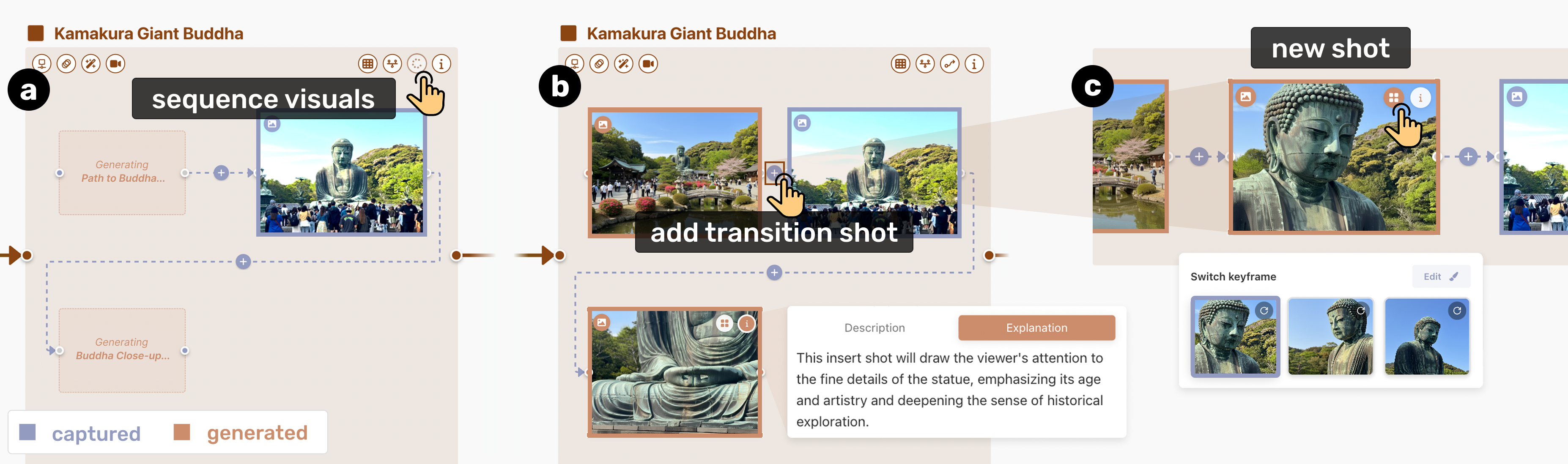}
    \caption{\textbf{Expanding a scene} by \textbf{(a)} sequencing visuals and suggesting \textcolor{AIColor}{new} content or \textbf{(b)} adding a transition shot between \textcolor{RealColor}{existing} footage. \textbf{(c)} Each generated shot comes with three options, and an explanation for how it strengthens the narrative.}
    \Description{Illustration of expanding a scene. (a) Sequencing visuals, (b) adding a transition shot, and (c) inserting a new generated shot with explanation of its narrative role. Generated content comes with multiple options for selection.}
    \label{fig:visual_sequence}
\end{figure*}
}

\newcommand{\figureControls}{
\begin{figure*}[ht]
    \centering
    \includegraphics[width=0.7\linewidth]{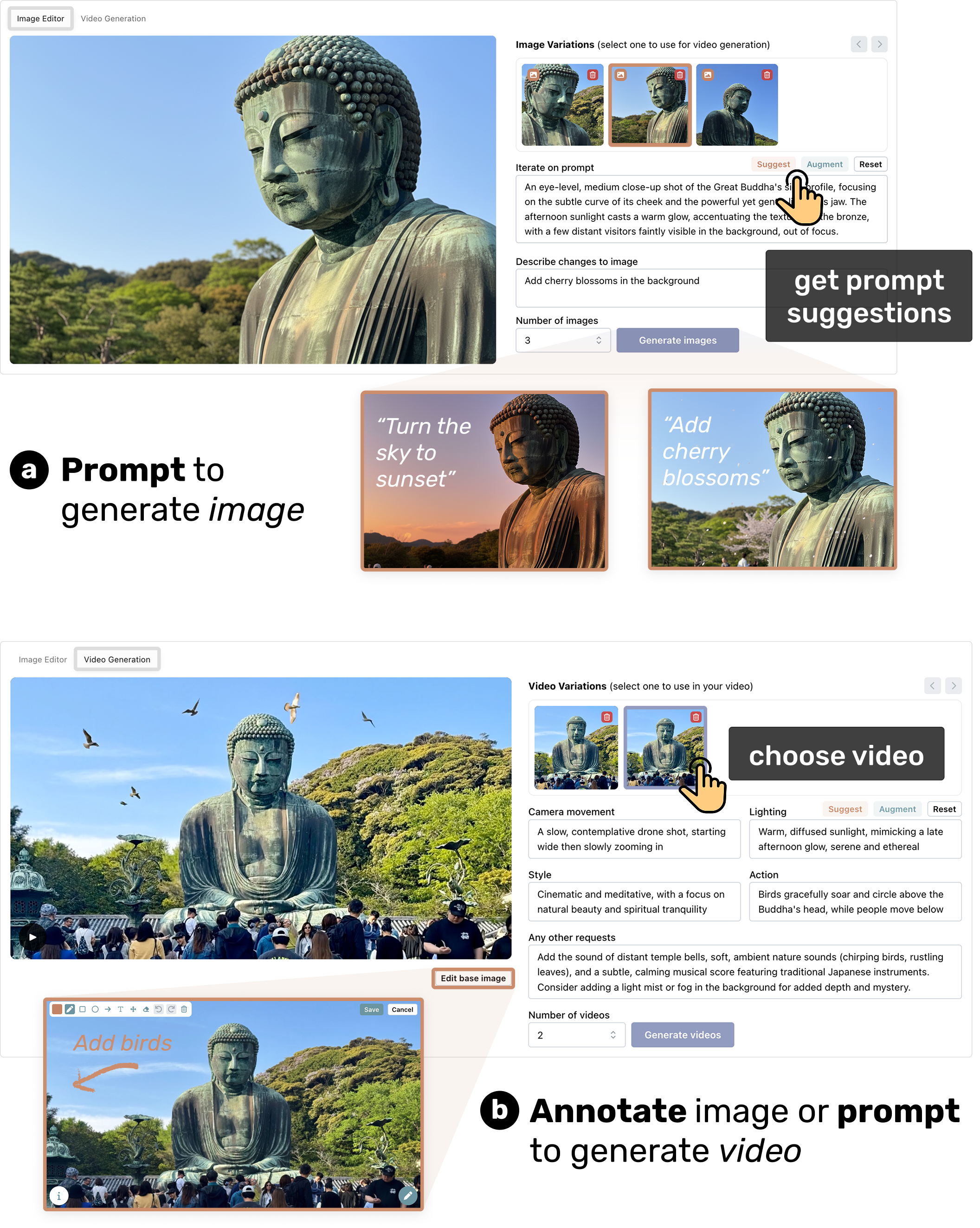}
    \caption{\textbf{Creative controls for refining generative outputs.} \textbf{(a)} Users can prompt to generate image variations in the \textit{image editor panel}. 
    Here, users can also ask the system for prompt suggestions. \textbf{(b)} Similarly, users can annotate images or write custom detailed prompts to bring them to life through generative video in the \textit{video generation panel}. To choose which clip to use in their video story, the user can simply select a variation in the gallery.}
    \Description{Screenshot showing creative controls for refining generative outputs. (a) Users can prompt to generate image variations, such as changing time of day or adding elements. (b) Users can annotate images or write prompts to generate video. Example shows adding cherry blossoms and birds to a scene with a statue.}
    \label{fig:controls}
\end{figure*}
}

\newcommand{\figureTimeline}{
\begin{figure}[t]
    \centering
    \includegraphics[width=\linewidth]{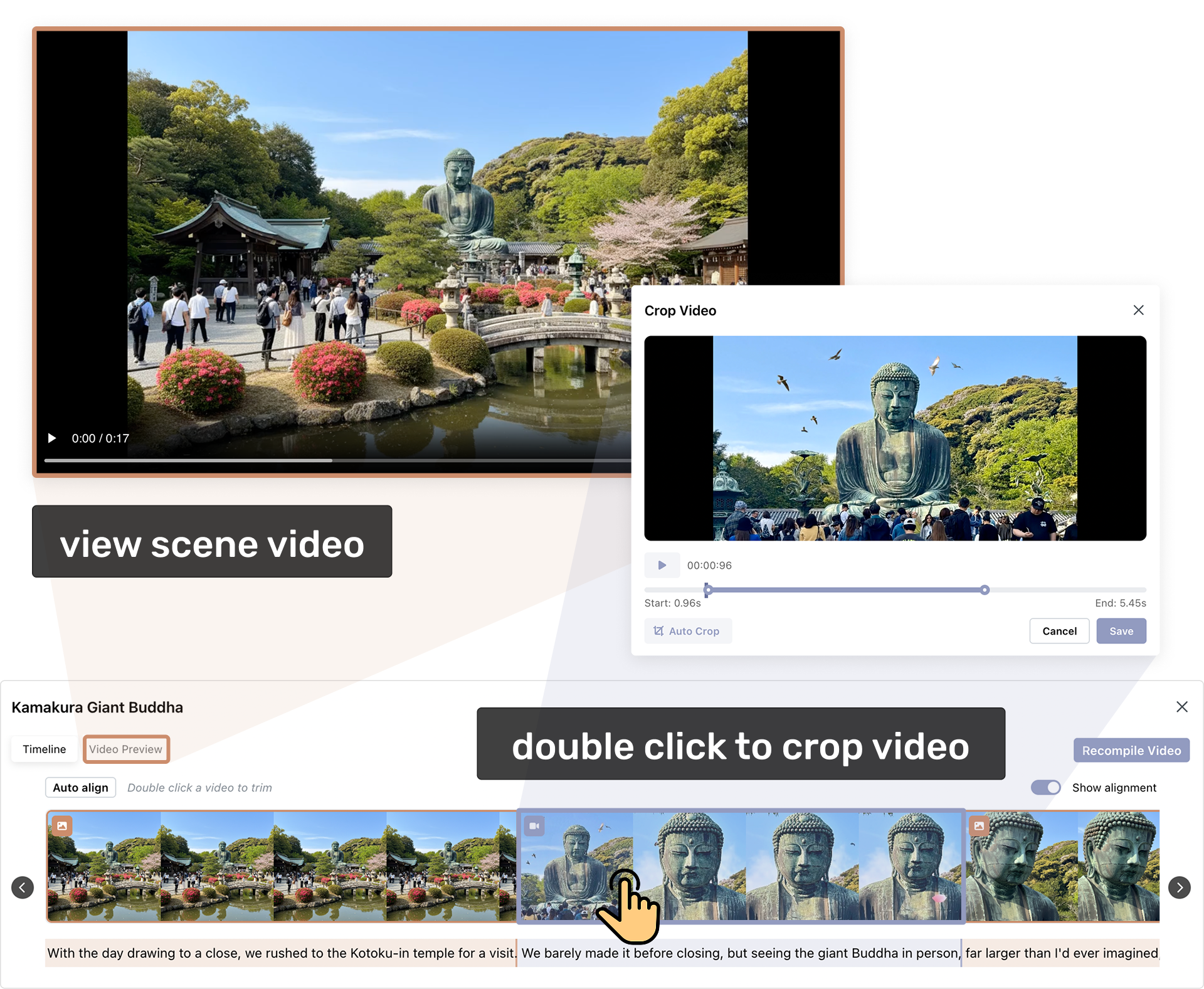}
    \caption{\textbf{The scene timeline} helps users align visuals to their script, so that narration will be synced automatically. Users can also crop clips and preview the scene video.}
    \Description{Screenshot of the scene timeline editor. Users can align visuals with the script, crop video clips, adjust timing, and preview the compiled scene. Example shows aligning narration with video of a temple garden and a Buddha statue.}
    \label{fig:timeline}
\end{figure}
}

\newcommand{\figureGenerativePipeline}{
\begin{figure*}[t]
    \centering
    \includegraphics[width=\linewidth]{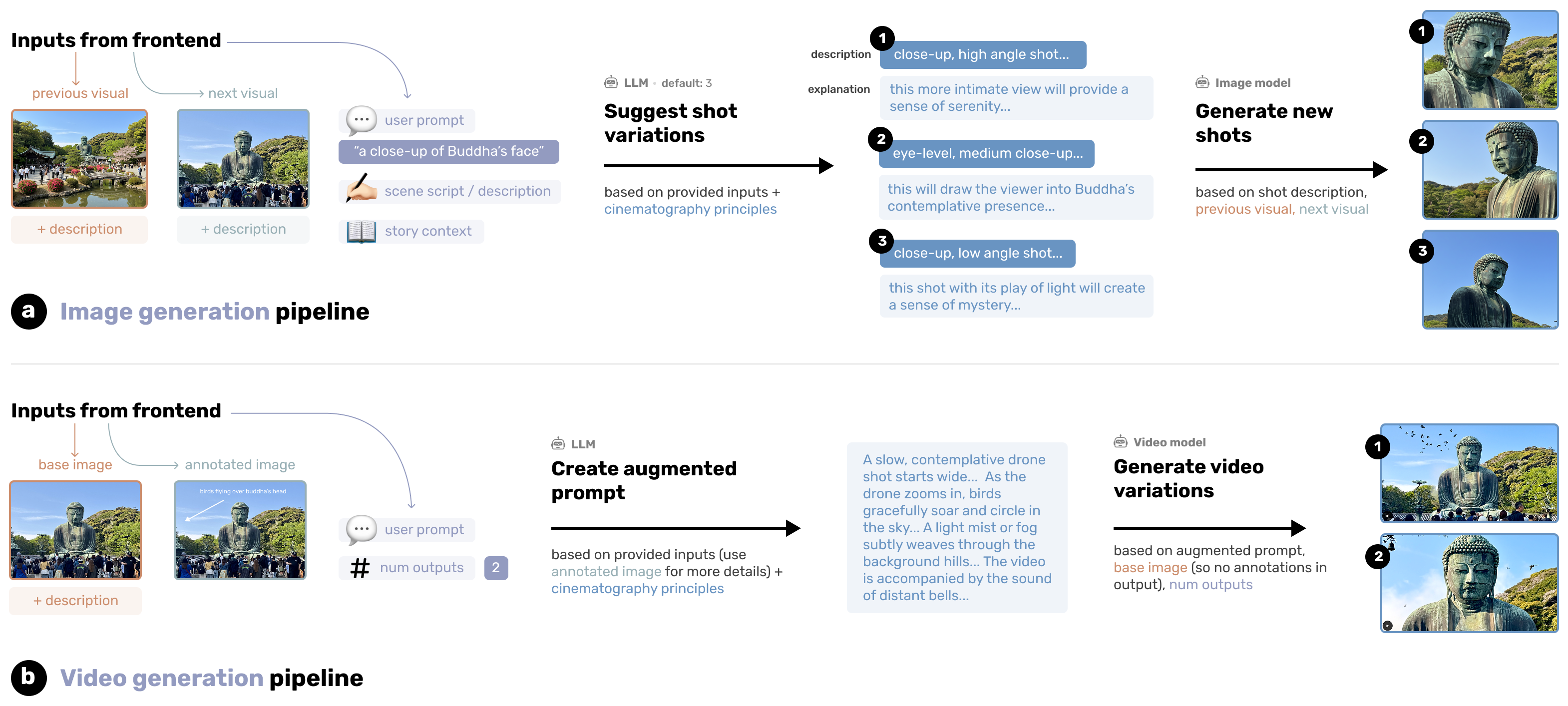}
    \caption{\textbf{Prompting pipeline for generating new contextual images and video variations.} \textbf{(a)} We generate contextualized image suggestions by drawing from neighboring visuals, along with scene / overall story context. \textbf{(b)} To generate a video variations, we create an augmented prompt from the base image and the user's annotations / prompt.}
    \Description{Pipeline diagram for generating new contextual images and videos. (a) Image generation pipeline: neighboring visuals, descriptions, and prompts are used to generate three image options with explanations. (b) Video generation pipeline: a base image, annotations, and prompts are used to create multiple video variants, with AI generating augmented prompts to guide the process.}
    \label{fig:generative_pipeline}
\end{figure*}
}

\newcommand{\figureSuggestionPipeline}{
\begin{figure*}[ht]
    \centering
    \includegraphics[width=\linewidth]{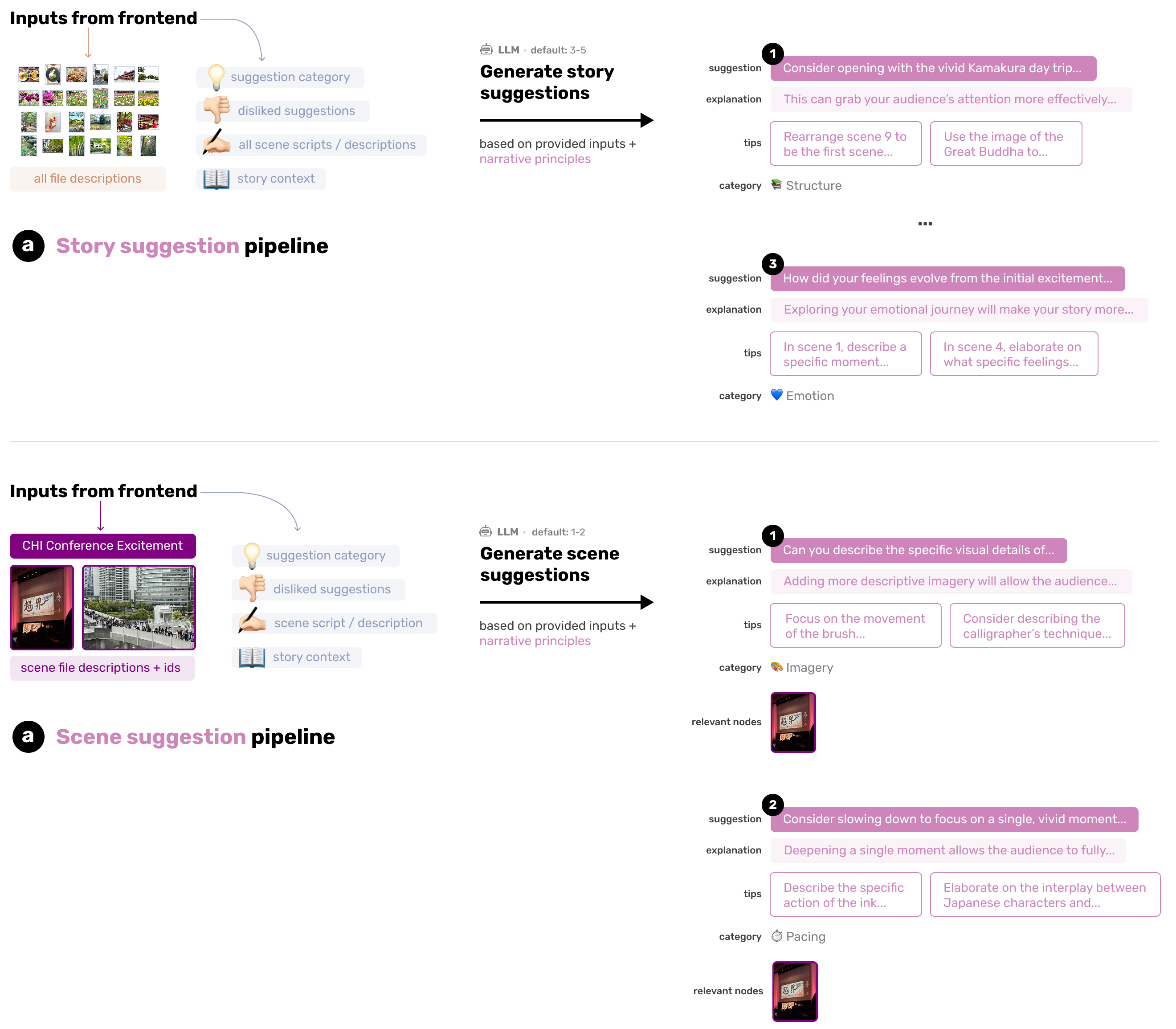}
    \caption{\textbf{Prompting pipeline for generating story-level and scene-level script suggestions.} \textbf{(a)} For \textbf{story suggestions,} we begin by passing in all the available file descriptions form the canvas, along with all scene scripts / descriptions, the overall story context, and the user's desired suggestion category and disliked suggestions (if available). Using this information (along with established narrative principles), we ask the LLM to generate 3-5 suggestions for enhancing the user's overall story. Each suggestion comes with an explanation of how it will make the script more compelling, along with 1-2 specific tips on how to address each suggestion. \textbf{(b)} For \textbf{scene suggestions,} we pass in similar inputs from the frontend, but focus on file descriptions and the script / description for the \textit{current scene.} Similar to story suggestions, we use this information (along with our narrative principles) to generate 1-2 suggestions for enhancing the user's script for this scene. For script suggestions, each output also includes the relevant node IDs for highlighting connections between scene's visual and text components (see \autoref{sec:interface}).}
    \Description{Pipeline diagram for generating story-level and scene-level script suggestions. (a) Story suggestions: AI analyzes available descriptions, scripts, and prompts to propose 3–5 suggestions with explanations and tips. (b) Scene suggestions: AI uses similar inputs but focuses on file descriptions and scene context, generating 1–2 refinements for improving narrative flow.}
    \label{fig:suggestion_pipeline}
\end{figure*}
}

\newcommand{\figureLikert}{
\begin{figure*}[t]
    \centering
    \includegraphics[width=\linewidth]{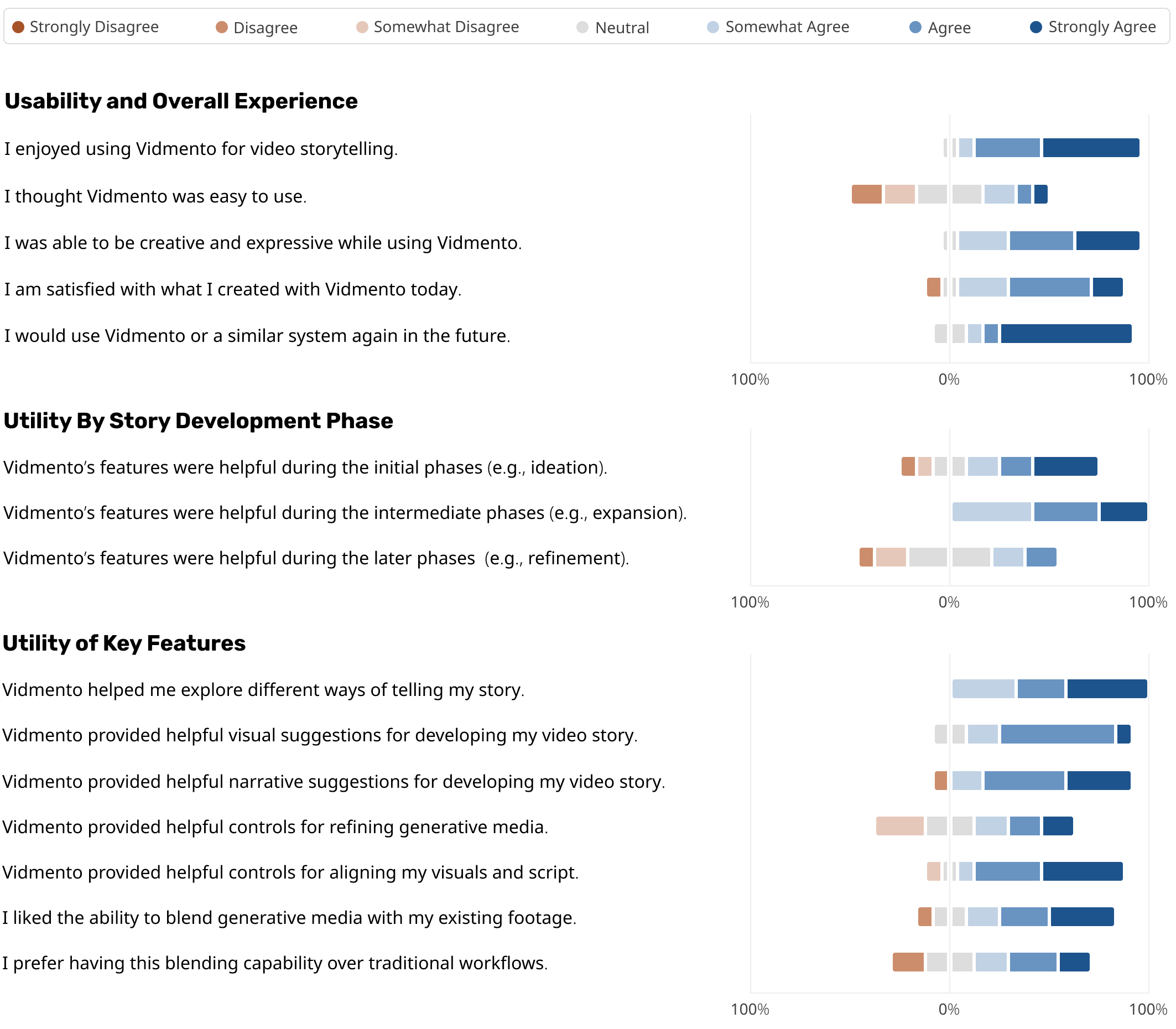}
    \caption{Participant ratings of the usability and utility of \system and its key features.}
    \Description{Bar chart showing participant ratings of Vidmento’s usability and utility. Items are grouped into three categories: overall experience, utility by story development phase, and utility of key features. Most participants agreed or strongly agreed that Vidmento was enjoyable, easy to use, and supported creativity. Utility was highest for ideation and intermediate phases. Participants especially valued blending generative and captured media, though some were neutral or disagreed about preferring it over traditional workflows.
}
    \label{fig:likert}
\end{figure*}
}

\newcommand{\figureCaseStudyOne}{
\begin{figure*}[t]
    \centering
    \includegraphics[width=\linewidth]{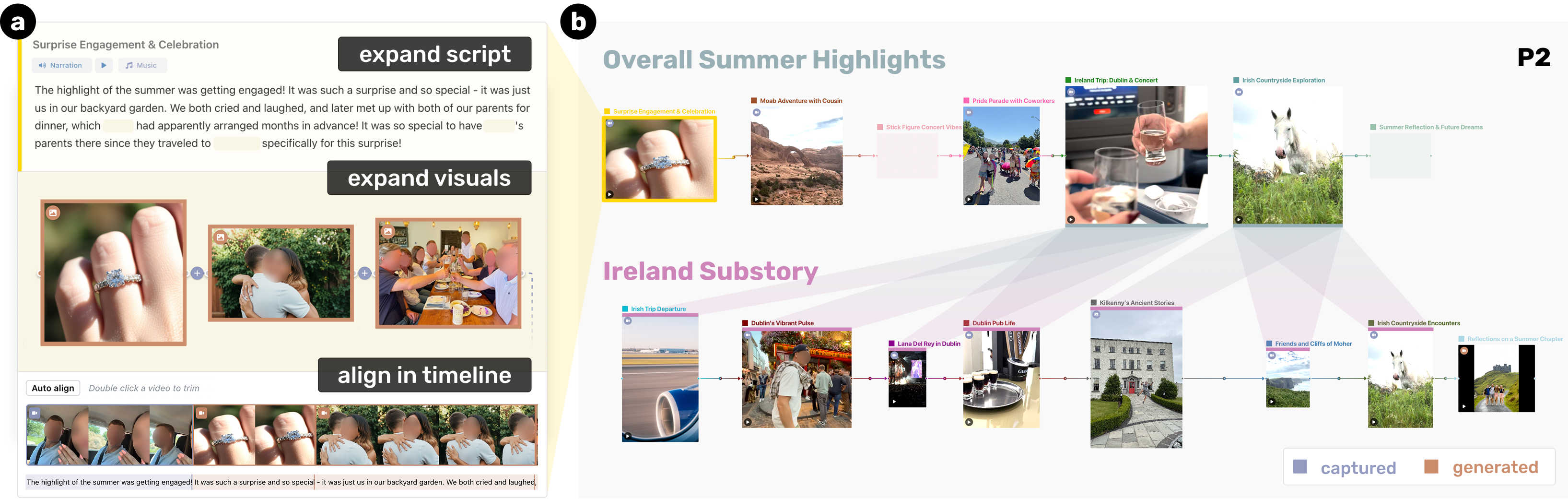}
    \caption{Showing how \textbf{P2} \textbf{(a)} expanded the script and visuals for the engagement scene in her video story and \textbf{(b)} explored a substory of her \textcolor{GreenColor}{summer highlights} focused on her \textcolor{PinkColor}{Ireland trip} using \system.}
    \Description{Example from participant P2. (a) Expanded the script and visuals for an engagement scene. (b) Explored a substory of summer highlights, focusing on an Ireland trip. Visual layout shows connected sequences with both captured and generated clips.}
    \label{fig:p2}
\end{figure*}
}

\newcommand{\figureCaseStudyTwo}{
\begin{figure*}[t]
    \centering
    \includegraphics[width=\linewidth]{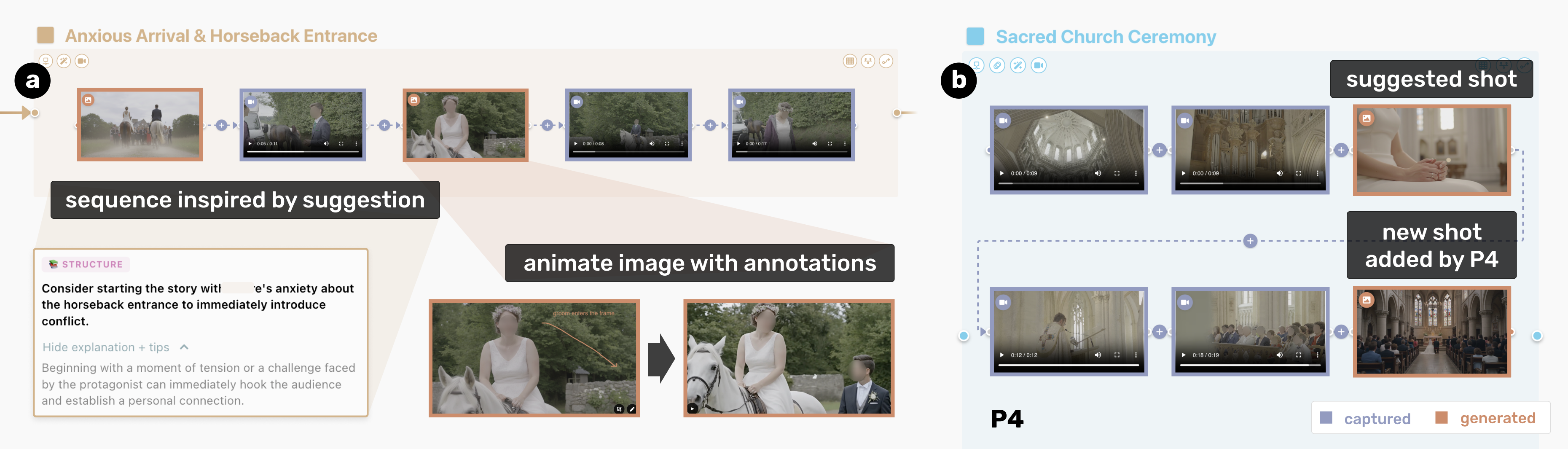}
    \caption{Showing how \textbf{P4} \textbf{(a)} expanded and animated the visuals for the wedding arrival scene in his video inspired by a story suggestion and \textbf{(b)} found new storytelling inspiration from \system's suggested visual sequence.}
    \Description{Example from participant P4. (a) Expanded and animated visuals for a wedding arrival scene using suggestions and annotations. (b) Found new storytelling inspiration by adding a suggested shot and generating new ones, creating a richer visual sequence.}
    \label{fig:p4}
\end{figure*}
}

\newcommand{\figureStoryboard}{
\begin{figure}[ht]
    \centering
    \includegraphics[width=0.6\linewidth]{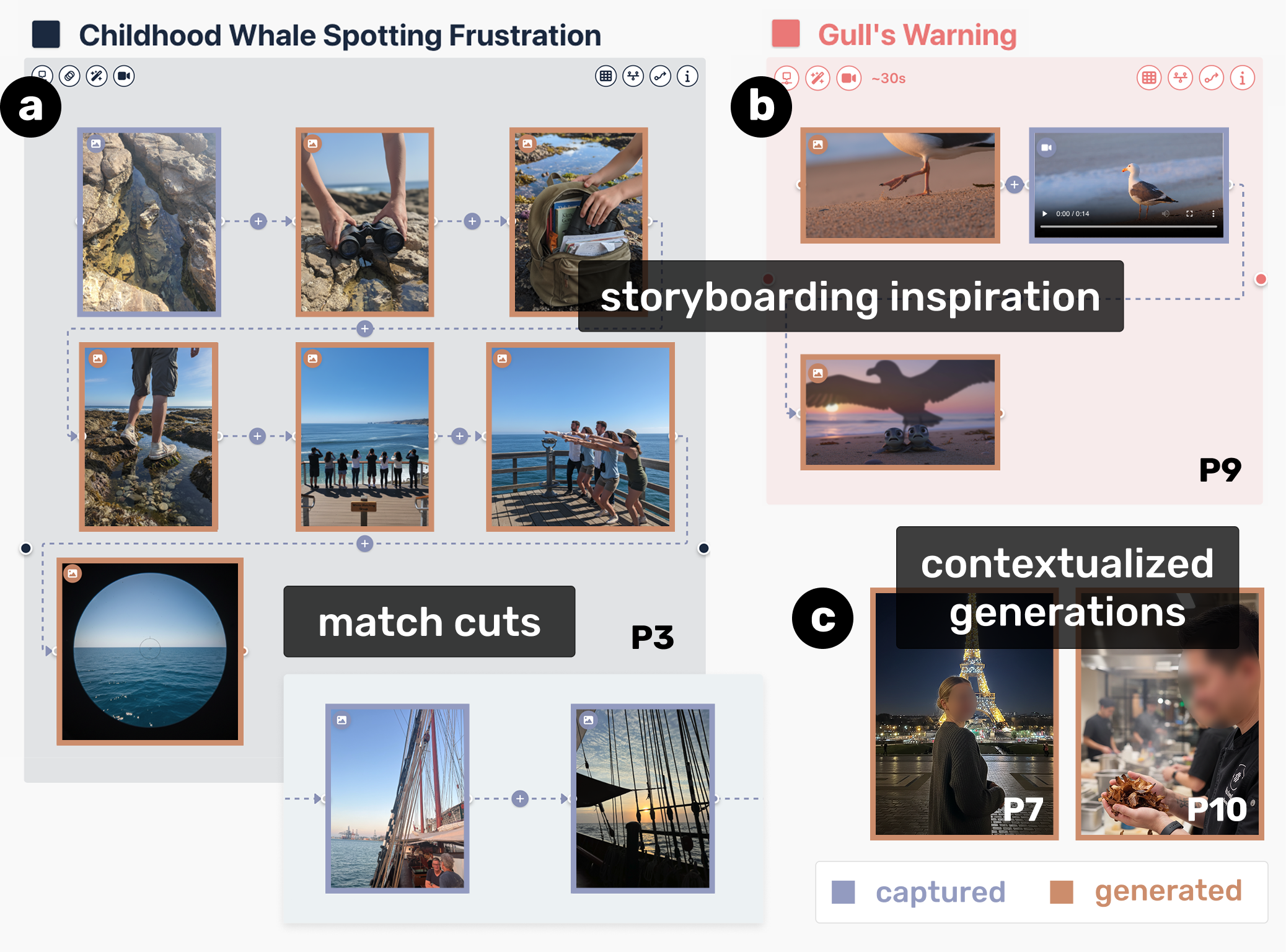}
    \caption{\textbf{Examples of successful visual expansions.} \textbf{(a)} P3 and \textbf{(b)} P9 used \system to generate new storytelling inspiration and storyboard their ideas. \textbf{(c)} P7 and P10 leveraged contextual generations to expand their stories.}
    \Description{Examples of successful visual expansions. (a) P3 used Vidmento to generate storyboard ideas through match cuts. (b) P9 generated shots to support a fictional narrative. (c) P7 and P10 used contextualized generations, where new media visually matched existing footage.}
    \label{fig:storyboard}
\end{figure}
}

\newcommand{\figureCreativeStorytelling}{
\begin{figure}[ht]
    \centering
    \includegraphics[width=0.65\linewidth]{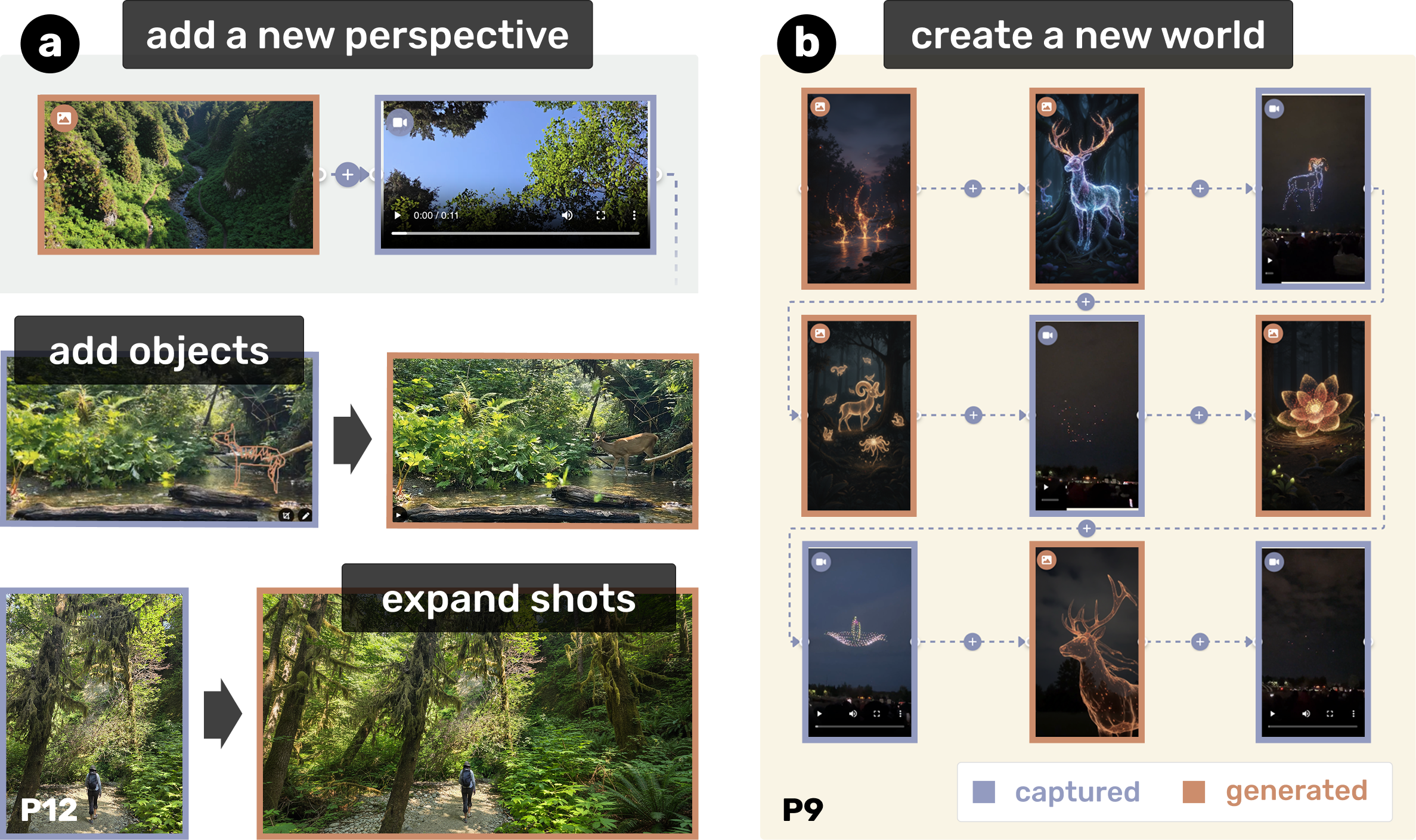}
    \caption{\textbf{Examples of bringing images to life with \system.} \textbf{(a)} P12 added new perspectives and objects to enhance his story, along with cinematic reframes of existing shots. \textbf{(b)} P9 leveraged his captured footage to imagine a fictional world inspired by the patronuses in \textit{Harry Potter}.}
    \Description{Examples of how creators used Vidmento’s generative features. (a) P12 added new perspectives, objects, and expanded shots to enhance captured footage. (b) P9 created a fictional world inspired by Harry Potter, blending captured and generated video.
}
    \label{fig:creative_storytelling}
\end{figure}
}

\newcommand{\figureStoryboardAndCreative}{
\begin{figure*}[ht]
\figureStoryboard
\hfill
\figureCreativeStorytelling
\end{figure*}
}

\newcommand{\figureDesignSpace}{
\begin{figure}[t]
    \centering
    \includegraphics[width=\linewidth]{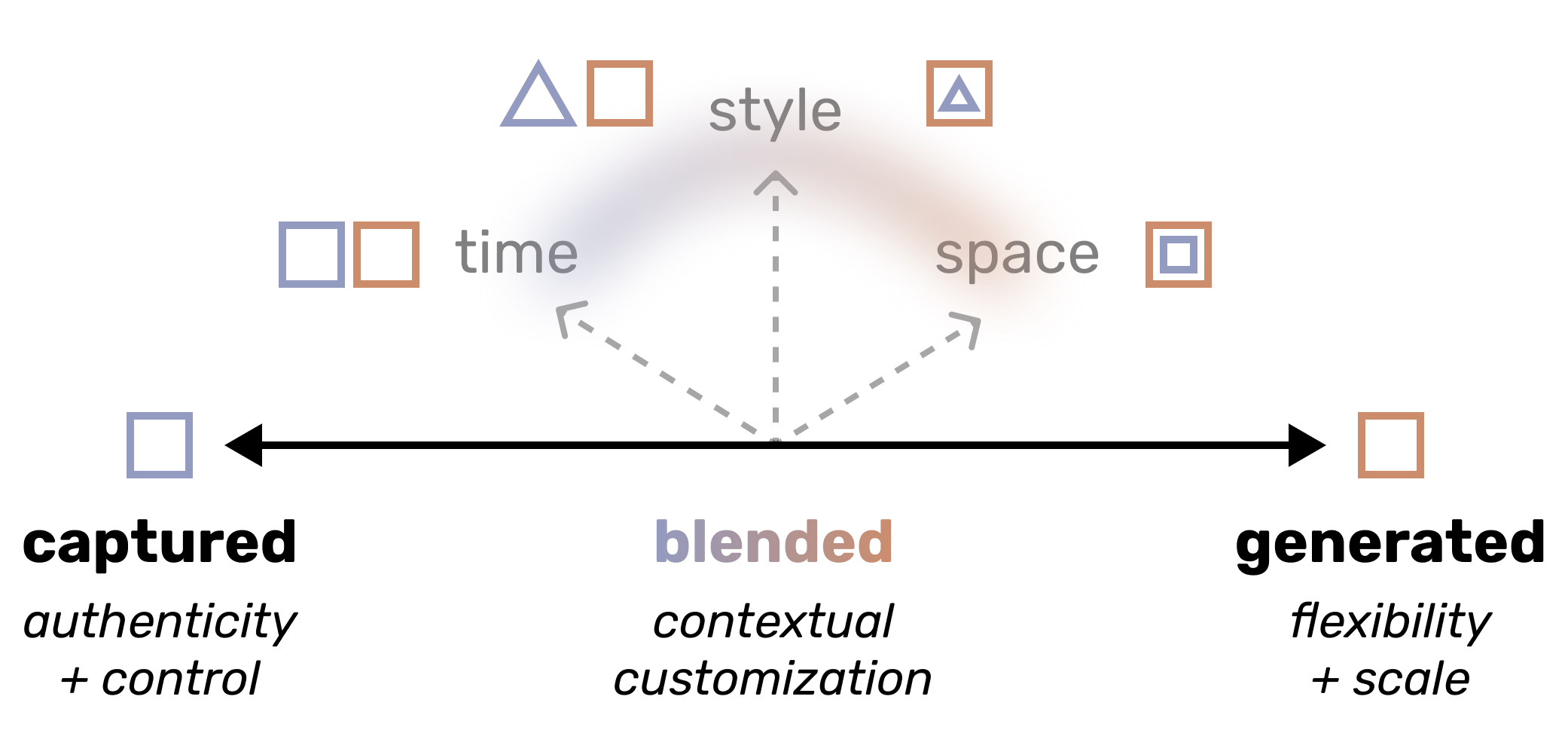}
    \caption{Positioning our \textbf{blended} video creation approach on the spectrum from fully captured to generative workflows. 
    We see opportunities to explore blending \textcolor{RealColor}{captured} and \textcolor{AIColor}{generated} video across time, space, and style. 
    Each shape represents a single shot.}
    \Description{Conceptual diagram positioning video workflows on a spectrum. On the left, captured footage represents authenticity and control. On the right, generative footage represents flexibility and scale. In between, blended workflows emphasize contextual customization. Shapes along the spectrum represent individual shots.}
    \label{fig:design_space}
\end{figure}
}

\newcommand{\tableFormPart}{
\begin{table*}[t]
{\sffamily
\begin{tabular}
{@{}p{0.5cm}p{3.5cm}p{4cm}p{3.5cm}p{4.9cm}@{}}
\textbf{ID} & \textbf{Role}             & \textbf{Video Editing Experience} & \textbf{Gen Video Experience} & \textbf{Video Story Topics}            \\ \midrule
F1          & University Student        & Novice (2 yrs)                    & No                                   & Yearly dumps, short films              \\
F2          & Director of Engineering   & Pro (30 yrs)                      & Yes (0.5 yrs)                        & Short films based on reality           \\
F3          & Freelancer                & Pro (6 years)                     & No                                   & Day in the life, short films           \\
F4          & University Student        & Intermediate (3 yrs)              & Yes (0.5 yrs)                        & Short films about personal stories \\
F5          & Strategic Dev. Manager    & Intermediate (16 yrs)             & Yes (1.5 yrs)                        & Travel and daily vlogs                 \\
F6          & Global Learning Architect & Intermediate (10 yrs)             & Yes (1 yr)                           & Family stories                         \\
F7          & Design Manager            & Pro (20 yrs)                      & Yes (1 yr)                           & Stories about kids, short films        \\
F8          & Professor                 & Novice (1 yr)                     & Yes (0.5 yrs)                        & Reflections on academic journey
\end{tabular}
}
\caption{Overview of our formative study participants.}
\Description{Table summarizing eight formative study participants (F1–F8). It lists each participant’s role, years of video editing experience (ranging from novice to professional), generative video experience (most with less than 1.5 years), and the topics of video stories they typically create (e.g., personal stories, vlogs, short films, family stories).}
\label{tab:form_part}
\end{table*}
}

\newcommand{\tableBooks}{
\begin{table*}[ht]
{\sffamily
\renewcommand{\arraystretch}{1.2}
\begin{tabular}{@{}p{11.5cm}p{4.9cm}r@{}}
\textbf{Title}              & \textbf{Author(s)} & \textbf{Year} \\ \midrule
Film Directing Shot by Shot: Visualizing From Concept to Screen~\cite{katz1991film} & Steven D. Katz     & 1991         \\ 
Story: Substance, Structure, Style, and the Principles of Screenwriting~\cite{mckee1997story}          & Robert McKee                  & 1997          \\
Save the Cat! The Last Book on Screenwriting That You'll Ever Need~\cite{synder2005save}              & Blake Snyder       & 2005          \\
The Anatomy of Story: 22 Steps to Becoming a Master Storyteller~\cite{truby2007anatomy}        & John Truby         & 2007          \\
Dialogue: The Art of Verbal Action for the Page, Stage, and Screen~\cite{mckee2016dialogue}       & Robert McKee                  & 2016          \\
Grammar of the Edit: Fourth Edition~\cite{bowen2018grammar}       & Christopher J. Bowen                 & 2018          \\
Character: The Art of Role and Cast Design for Page, Stage, and Screen~\cite{mckee2021character}      & Robert McKee                  & 2021          \\
Action: The Art of Excitement for Screen, Page, and Game~\cite{mckee2022action}         & Robert McKee, Bassim El-Wakil & 2022          \\ 
The Anatomy of Genres: How Story Forms Explain the Way the World Works~\cite{truby2022anatomy}       & John Truby         & 2022 
\end{tabular}
}
\caption{Overview of the \numbooks books we read to learn more about storytelling and filmmaking.}
\Description{Table listing nine filmmaking and storytelling books used to inform Vidmento’s design. Each row includes the book title, author(s), and year. Examples include Film Directing Shot by Shot (1991, Katz), Story: Substance, Structure, Style, and the Principles of Screenwriting (1997, McKee), and The Anatomy of Genres (2022, Truby).}
\label{tab:books}
\end{table*}
}

\newcommand{\tableCategories}{
\begin{table*}[ht]
{\sffamily
\renewcommand{\arraystretch}{1.2}
\begin{tabular}{@{}p{2cm}l@{}}
\textbf{Category} & \textbf{Description}                                                                                    \\ \midrule
Structure                   & Story chronology or organization                                                                        \\
Plot                        & Story progression or narrative flow (\eg conflicts, rising action, falling action, etc.) \\
Imagery                     & Visual details or descriptive elements                                                                  \\
Character                   & Character development or motivations                                                                    \\
Dialogue                    & Conversation or speech patterns                                                                         \\
Pacing                      & Story timing, rhythm, or tempo                                                                          \\
Emotion                     & Feelings, mood, or emotional impact                                                                     \\
Setting                     & Story location, atmosphere, or environment                                                              \\
Theme                       & Underlying messages or meaning in the story                                                             \\
Other                       & Other aspects not covered by the listed categories                      
\end{tabular}
}
\caption{Overview of the narrative categories we used for script suggestions.}
\Description{Table outlining narrative categories used for script suggestions. Ten categories are listed with descriptions: structure (story chronology), plot (progression and flow), imagery (visual details), character (development or motivations), dialogue (conversations), pacing (timing or tempo), emotion (feelings and mood), setting (location and atmosphere), theme (underlying meaning), and “other” (aspects not covered by listed categories).}
\label{tab:categories}
\end{table*}
}

\newcommand{\tableStudyPart}{
\begin{table*}[t]
{\sffamily
\begin{tabular}{@{}p{0.5cm}p{3.25cm}p{4cm}p{3.5cm}p{5.15cm}@{}}
\textbf{ID} & \textbf{Role}                      & \textbf{Video Editing Experience} & \textbf{Gen Video Experience} & \textbf{Video Story Topic}            \\ \midrule
P1                   & Designer                           & Pro (25+ yrs)                     & Yes (2 yrs)                          & Event reel for social media           \\
P2                   & Content Strategist                 & Novice (10 yrs)                   & No                                   & Summer highlights / engagement video  \\
P3                   & PhD Student                    & Intermediate (5 years)            & Yes (1.5 yrs)                        & Personal essay about sailing          \\
P4                   & PhD Student                        & Pro (9 yrs)                       & Yes (1 yr)                           & Wedding video                         \\
P5                   & Enterprise Architect               & Novice (0.5 yrs)                  & Yes (1 yr)                           & Fantasy story based on drone art show \\
P6                   & University Student & Intermediate (7 yrs)              & Yes (1 yr)                           & Internship highlights                 \\
P7                   & Social Media Manager               & Intermediate (4 yrs)              & Yes (1 yr)                           & Personal essay about moving abroad    \\
P8                   & Product Manager                    & Novice (2 yrs)                    & No                                   & Day in my life - work event          \\
P9                   & Product Manager                    & Intermediate (5 yrs)              & Yes (0.5 yrs)                        & Fictional story from stock footage    \\
P10                  & Learn Content Lead                 & Intermediate (8 yrs)              & Yes (1 yr)                           & Gardening / cooking demo              \\
P11                  & Software Engineer                  & Novice (1 yr)                     & No                                   & Daughter growing up story             \\
P12                  & Solution Architect                 & Pro (10 yrs)                      & No                                   & Elevated version of hiking trip 
\end{tabular}
}
\caption{Overview of our user study participants.}
\Description{Table summarizing twelve user study participants (P1–P12). Columns list role, years of video editing experience (from novice to professional), generative video experience (ranging from none to 2 years), and their video story topics (e.g., event reels, wedding video, internship highlights, fictional stories, cooking demos, hiking trips).}
\label{tab:study_participants}
\end{table*}
}

\newcommand{\tableFeatures}{
\begin{table*}[ht]
{\sffamily
\renewcommand{\arraystretch}{1.5}
\begin{tabular}{@{}p{4cm}p{7cm}p{6.1cm}@{}}
\textbf{Feature}             & \textbf{Inputs}                                                                                                        & \textbf{Outputs}                                                                    \\ \midrule
 \llm \textit{Generate shot description}    & Shot (image/video), \gray{story context}                                                                                      & Shot description                                                                    \\
 \llm \textit{Generate scene description}    & Shot descriptions in scene, \gray{story context}                                                                              & New scene title, description, color                                                 \\
 \llm \textit{Group all shots into scenes}   & All shot ids + descriptions, \gray{story context}                                                                             & New scenes (title, description, color, shot ids)                                    \\
 \llm \textit{Sequence all scenes}           & All scene data (name, description), \gray{story context}                                                                      & Ordered scene data                                                                  \\
 \llm \textit{Add contextual scene}      & Previous scene title + description, next scene title + description, \gray{story context}                                      & New scene title, description, color                                                 \\
 \llm \textit{Create story variation}      & All shot ids + descriptions, \gray{story context, user prompt}                                                                & Ordered new scenes (title, description, color, shot ids)                            \\
 \llm \textit{Compare story variations}     & Name, ordered scene data (description, script) for each variation                                                      & Compare data (summary + list of pros and cons for each variation)                   \\
 \llm \textit{Generate story suggestions}             & All shot descriptions, \gray{all scene scripts + descriptions, story context, disliked suggestions, suggestion category}      & Suggestions with explanation + tips                                                 \\
 \llm \textit{Generate scene suggestions}             & Shot descriptions + ids in scene, \gray{scene script} + description, \gray{story context, disliked suggestions, suggestion category} & Suggestions with explanation + tips + relevant shot ids                             \\
 \llm \textit{Sync notes to script}    & Story context, scene ids + descriptions                  & Segmented data (mapping scenes to script segments)                                  \\
  \llm \textit{Refine text}    & Original text, \gray{user prompt}                  & Three refined text options                                  \\
  \llm \textit{Sequence visuals in scene}    & Shot ids + descriptions in scene, scene description + \gray{script, story context}                                              & Ordered shot data (including new shot suggestions)                                  \\
\llm \image \textit{Add contextual shot}     & Previous shot + description, next shot + description, \gray{scene script + description, story context, user prompt}           & Three shot variants with description + explanation                                  \\
 \llm \image \textit{Generate image variations}     & Base image + description, \gray{previous shot, next shot}, $n$ (number of outputs), \gray{user prompt}                                 & $n$ image variants with description + explanation                                     \\
\llm \video \textit{Generate video variations}    & Base image + description, \gray{annotated image}, $n$ (number of outputs), \gray{user prompt}                                          & $n$ video variants    \\
\narrate \textit{Generate narration}  & Scene script 
& Generated narration \\
\llm \music \textit{Generate music} & \gray{Scene script} + description, \gray{duration, $n$ (number of outputs), user prompt}
& Generated music \\
 \llm \textit{Auto align visuals to script} & Scene script, shot ids + descriptions                                                                                  & Ordered list of script correspondences (shot-text mappings)                         \\
\llm \textit{Compile scene video}           & Scene data (ordered shots, \gray{narration, music, script correspondences})                                                           & Compiled scene video (with auto-timed shots based on correspondences) 
\end{tabular}
}
\caption{Overview of our AI-powered system features. Inputs highlighted in \gray{gray} are optional. Colored dots indicate the model used (\llm = multimodal LLM, \image = image, \video = video, \narrate = audio, \music = music). }
\Description{Table summarizing Vidmento’s AI-powered system features. Columns list feature, inputs, and outputs. Features include generating shot and scene descriptions, grouping shots into scenes, sequencing, adding contextual scenes, creating story variations, comparing versions, generating or refining visuals, and aligning visuals to scripts. Icons indicate whether the feature uses multimodal models (LLM, image, video, or audio).}
\label{tab:system_features}
\end{table*}
}

\begin{abstract}
Video storytelling is often constrained by available material, limiting creative expression and leaving undesired narrative gaps. 
Generative video offers a new way to address these limitations by augmenting captured media with tailored visuals. 
To explore this potential, we interviewed eight video creators to identify opportunities and challenges in integrating generative video into their workflows. 
Building on these insights and established filmmaking principles, we developed \system, a tool for authoring hybrid video stories that combine captured and generated media through context-aware expansion.
\system surfaces opportunities for story development, generates clips that blend stylistically and narratively with surrounding media, and provides controls for refinement.
In a study with 12 creators, \system supported narrative development and exploration by systematically expanding initial materials with generative media, enabling expressive video storytelling aligned with creative intent.
We highlight how creators bridge story gaps with generative content and where they find this blending capability most valuable.
\end{abstract}



\figureTeaser

\maketitle

\section{Introduction}

Storytelling often begins with a small set of ideas that are gradually expanded into a complete narrative~\cite{truby2007anatomy, synder2005save}. 
In writing, this expansion is flexible: authors can freely add, revise, or remove words until the story achieves the desired shape~\cite{mckee1997story}. 
Video storytelling, by contrast, imposes stricter constraints once shooting begins. 
While trimming footage is simple, adding new material to fill undesired narrative gaps hinges on what was captured during filming~\cite{katz1991film, bowen2018grammar}. 
To guard against missing content, professionals often overshoot and rely on pick-up shots~\cite{miller2013script}, while everyday creators may only realize later that essential footage was never recorded. 
Stock media can sometimes bridge these gaps, but rarely aligns with a creator's vision or video context~\cite{sealmedia2024archetypes, allinmotion2023stock, skillman2025pros}. 

Generative video offers a promising medium to address these challenges by allowing creators to craft tailored clips on demand~\cite{aldausari2022video}, with recent models such as Veo3~\cite{veo3} demonstrating near production-level quality.
While fully synthetic video production is one possibility~\cite{googleflow, skyreels,stern2025this,toonkel2025open}, we focus on how generative video can augment captured materials.
Commercial tools now include features like ``Generative Extend''~\cite{adobegenextend, wang2025boundary}, but these elongate individual clips in isolation, without considering the larger narrative.
We see a deeper opportunity in contextually \textit{blending} the fixed realities of captured content with the open possibilities of generation to expand incomplete narratives into cohesive stories.

To probe this design space, we conducted formative interviews with eight video creators.
Participants often struggled to craft compelling narratives or explore alternatives when working with fixed footage.
As a result, they viewed generative video as a way to expand stories beyond what was recorded, but also raised concerns about maintaining continuity and creative autonomy.
This tension highlights the need for hybrid video creation approaches that bridge the grounded nature of captured media with the expansive potential of generation.

Motivated by these findings and an analysis of nine filmmaking books, we propose \textbf{generative expansion}, a design framework that articulates how generative video can contextually extend and interweave with captured footage for hybrid video storytelling.
Generative expansion comprises four design goals grounded in established storytelling principles: \textit{(1) identifying story ideas} that can be developed from the user's media, \textit{(2) highlighting opportunities} where new material could expand the story, \textit{(3) generating clips} that blend stylistically and narratively with existing footage, and \textit{(4) providing controls} that let users refine generations and preserve creative authorship when blending media.

We instantiate this framework in \system, an AI-assisted authoring tool that integrates captured and generated footage within a unified storytelling environment (\autoref{fig:teaser}).
\system is tailored for \textit{narrated video stories}, a widely-used genre in which voiceover narration and visuals work together to shape narrative meaning~\cite{truong2016quickcut,lu2023show}.
To support varied workflows, our system provides affordances for both high-level exploration and precise editing through a linked canvas and script editor. 
The canvas functions as a visual storyline where users can directly sequence media, visualize narrative gaps, and generate contextualized content to fill them.
It combines the flexibility of a freeform workspace and structure of an editing timeline to create a \textit{semi-structured} environment for video creation.
\system also draws from narrative theory and cinematography to provide Socratic-style suggestions for expanding voiceover scripts~\cite{gmeiner2025exploring,elder1998role}.

In a user study with 12 video creators, \system enabled expressive video storytelling by scaffolding the exploration and expansion of story possibilities grounded in their existing footage.
Creators emphasized that our system helped them engage with generative video in ways that amplified -- rather than diminished -- their creative voice, transforming initial ideas and raw media into compelling narratives that would be difficult to achieve with existing workflows. 
We discuss use cases and limitations for blending captured and generative visuals, outlining future opportunities for hybrid video storytelling.
To summarize, our contributions include: 
\begin{itemize}
    \item \textbf{Generative expansion, a hybrid video creation framework} that integrates generative and captured media to craft compelling, contextually-grounded video stories.
    \item \textbf{The design and implementation of \system,} a unified authoring tool that instantiates generative expansion to help users visualize, explore, and shape video storylines.
    \item \textbf{Findings from formative interviews and a user study} that reveal key opportunities and challenges in crafting blended video stories, laying foundations for an emerging video creation paradigm.
\end{itemize}

\section{Related Work}

We build on research in generative video creation, LLM-assisted video editing and story writing, and canvas-based interfaces for creative authoring.
To our knowledge, this is the first work to articulate a design framework and system that combines captured and generative video within a unified storytelling environment.

\subsection{Creating With Generative Video}
Generative video is rapidly emerging as a new creative paradigm~\cite{aldausari2022video}, with generated clips appearing in production work such as ads~\cite{horvath2024coca} and short films~\cite{stern2025this,toonkel2025open}. 
Most workflows use text-to-video models~\cite{singer2022make, wang2023modelscope, yang2024cogvideox, wang2025lavie, Yin_2025_CVPR} such as Veo3~\cite{veo3} and Sora~\cite{sora}, where creators can specify detailed prompts~\cite{veoprompt}, an initial frame~\cite{jiang2024videobooth,Wu_2023_ICCV}, or spatial annotations~\cite{spatialprompt} to control shot composition, motion, and continuity.  
These advances have enabled authoring tools like Google Flow~\cite{googleflow} and fully automated filmmaking pipelines such as FilmAgent~\cite{xu2025filmagent} and SkyReels~\cite{skyreels}.
However, existing systems treat generation as an isolated process rather than a means to augment footage or partially complete stories.
\system instead positions generative video as a complementary material, allowing users to contextually blend captured and generated clips within one narrative workflow.

\subsection{LLM-Assisted Video Creation}
Video creation often requires extensive planning, editing, and expertise.
Various algorithms and tools have been developed to streamline this process~\cite{truong2016quickcut, 2017leakeComp, leake2020generating, xia2020crosscast, casares2002simplifying, cattelan2008watch, zhang2016snapvideo, write-a-video, lin2022videomap, 10.1145/3544548.3581494,pavel2016vidcrit}.
A recent trend is using (multimodal) LLMs to analyze materials~\cite{wang2024lave}, shape stories~\cite{wang2024reelframerhumanaicocreationnewstovideo}, sequence clips~\cite{wang2024lave, marcelo2025editduet}, and extract highlights~\cite{wang2024podreelshumanaicocreationvideo, Barua_2025}. 
LLMs also introduce novel interactive editing paradigms.
For example, ExpressEdit~\cite{tilekbay2024expressedit} integrates text and sketching, while LAVE~\cite{wang2024lave} provides agentic editing assistance and VideoDiff~\cite{huh2025videodiff} supports side-by-side comparison of edited variants.
These systems focus primarily on editing existing footage.
In contrast, we leverage LLMs to identify narrative gaps and propose new generative clips to bridge and expand the story.

\subsection{Writing Stories With LLMs}
LLMs are widely used for story writing assistance~\cite{Lee_2024, shakeri2021saga, 10.1145/3706598.3714119}.
Systems such as Wordcraft~\cite{yuan2022wordcraft} and CoAuthor~\cite{Lee_2022} embed LLM-generated suggestions into text editors, while Talebrush \cite{chung2022talebrush} explores sketch-based story generation. 
While LLMs can produce full, coherent stories (\eg screenplays in Dramatron~\cite{mirowski2023co}), our formative study participants preferred drafting their own video scripts to preserve narrative intent, in line with prior work~\cite{yeh2024ghostwriter}. 
Accordingly, \system provides Socratic-style prompts~\cite{gmeiner2025exploring,elder1998role} to support reflection and iteration rather than automated story generation.

\subsection{Canvas-Based Interfaces for Creative Authoring}
Creative workflows are increasingly supported by canvas-based interfaces that organize ideas spatially rather than sequentially~\cite{Chung_2025, comfyui, amin2025promptcanvascomposablepromptingworkspaces, gonzalez2024collaborative,masson2024visual,qin2025toward}.
Examples include PatchView~\cite{chung2024patchview} for story worldbuilding, Luminate~\cite{suh2024luminate} for design space exploration during writing, and Promptify~\cite{brade23promptify} for image prompt iteration. 
Superstudio~\cite{superstudio}, FLORA~\cite{flora}, Midjourney Patchwork~\cite{midjourneypatchwork}, and Firefly Boards~\cite{fireflyboards} extend canvases to multimodal authoring.
Tools such as VISAR~\cite{visar2023zhang} and VideOrigami~\cite{cao2025compositional} link canvases with sequential UIs (\eg text or timeline editor) to combine the benefits of both -- a paradigm we build on.
These systems position the canvas as a place for freeform ideation, but do not explore how spatial layout can guide narrative flow.
We design our canvas as a visual storyline where creators can directly sequence and edit scenes -- in addition to visualizing story gaps and ideas to fill them.

\section{Formative Study}\label{sec:form_study}
To probe the design space of blended video storytelling that combines captured and generated media, we conducted formative interviews to investigate (1) creators' \textit{current practices for developing videos from an initial set of captured materials} and (2) \textit{opportunities and challenges of integrating generative footage into those workflows}. 
As practices differ across creators and genres, our goal was not to characterize a single process, but to surface this variation and identify where generative video could meaningfully extend capture-only workflows and address their limitations.

\tableFormPart
\boldpar{Participants} We interviewed eight video creators (\textbf{F1–8}) recruited through internal mailing lists and snowball sampling: four full-time employees at \location, two university students, one professor, and one freelancer (\autoref{tab:form_part}). 
Three self-identified as professional video editors (\textit{avg:} 18.7 years of experience), three as intermediate (\textit{avg:} 9.7 years), and two as novices (\textit{avg:} 1.5 years). 
Six had 0.5–1.5 years of experience with generative video models (\eg Sora, Firefly), while two had none, included to broaden perspectives. 
Participants' work ranged from travel vlogs and family narratives to short films.

\boldpar{Protocol} Each one-hour, semi-structured interview began with participants sharing 1–2 videos created from their own materials (using generative video when applicable) and describing their storytelling workflows. 
Those with prior experience reflected on how generative tools supported or challenged their process, while those without speculated on potential uses. 
Participants received \$40 in compensation. 

\subsection{How Do Creators Build Stories From Captured Materials?} \label{sec:form_study_sub1}
When working solely with captured media, all editors began with a core idea or story premise,
often anchored in specific artifacts such as photos, letters, and recordings ($n=5$).
F8 described how she starts with \myquote{what's a feeling or core experience I had and how can I make it resonate on a screen.}
From there, creators either developed their stories in a linear text editor ($n=3$) or a freeform canvas ($n=4$): \myquote{My brain works spatially and I like to cluster things so I hate working in documents} (F6).
For \myquote{honing the pieces after you know the general structure,} most used a timeline editor (F1).
Overall, creators valued a \textbf{flexible, unified workspace} that supports iteration and does not \myquote{force me into one particular workflow} (F7). 
F8 envisioned a system that allows \myquote{collecting existing photos [and] creating digital assets in one spot... that would be hugely beneficial.}

\boldpar{Challenges in Capture-Only Workflows}
Three creators struggled with \textbf{finding the right order and pacing} when \myquote{weaving footage together} (F8), which they viewed as important for crafting a compelling narrative, echoing prior work~\cite{cao2025compositional,cao2023dataparticles}.
As F5 said, \myquote{I'm always thinking about how [to] keep things moving but not too jerky. That's the part I struggle with.}
Two cited \textbf{limited footage and filming resources} as constraining the set of stories they could tell ($n=2$): \myquote{I'd have to construct a set I can't afford, or get permission to film in a place I can't} (F2).

\subsection{Opportunities For Augmenting Video Storytelling With Generative Video}
\label{sec:form_study_opp}
Creators viewed generative video as a way to address creative bottlenecks with captured footage, identifying three key opportunities.

\boldpar{O1: Supporting story construction and exploration}
Three participants thought generative video could help connect artifacts into cohesive stories: \myquote{[AI] helps me see things in new combinations. And that's what stories are.} (F8).
Four imagined using it to fill content gaps -- similar to B-roll / stock footage -- or expand existing clips ($n=2$), \eg by \myquote{generating new perspectives or shots from different angles} (F3). 
Three emphasized the ability to quickly visualize ideas: \myquote{Because you can generate variations so fast, if you want to see how something would look, that's really useful} (F3). 

\boldpar{O2: Enabling previously impossible storytelling possibilities}
Seven saw potential for bringing untold or undocumented stories to life by overcoming the limits of available footage.
F4 reflected, \myquote{My grandma immigrated to Brazil, but we don't have any record of that... maybe the AI could generate videos if I gave it some photos.} 
Similarly, F6 imagined reconstructing community histories: \myquote{If you could show how the neighborhood actually looked back in the 60s, that would be really cool.}
Generative video can also transcend filmmaking constraints such as inaccessible sets (F2).

\boldpar{O3: Reimagining narratives through expressive reinterpretation}
Six envisioned using generative video to \myquote{express things in interesting and new ways} (F8).
F7 described depicting her family as octopuses to document a day in her life as a mom.
Three others wanted to augment tangible artifacts and experiences.
F5 thought it \myquote{would be cool if you could draw something and then it would bring it to life,} and F1 imagined using AI to animate her story about learning piano.

\subsection{Challenges With Blending Generative Video With Captured Footage}
\label{sec:form_study_cha}
However, participants also emphasized frictions introduced by generative video.

\boldpar{C1: Integration issues}
Creators reported inconsistencies (\eg style, motion) between generated and captured footage ($n=5$)~\cite{qu2024exploring,wang2025boundary,xiao2025videoauteur}, limited control when prompting ($n=7$)~\cite{rozo2025prompt,geng2025motion}, and slow iteration times ($n=6$)~\cite{dedhia2025generating,chen2024adaptivegenbackend,yin2025causvid}.
F8 shared how working with AI \myquote{is like a slot machine... You might get something that you're happy with,} but other times, \myquote{you just get the same hard cut over and over [that] doesn't work and that's very frustrating.}

\boldpar{C2: Authenticity concerns}
Three worried about misrepresenting lived experiences or losing authenticity when adding generative elements to videos~\cite{polimetla2025paradigm,shi2023understanding}.
F6 explained, \myquote{If [AI] generated something false that's generally the feel of this mountain range, but not the mountain range [and] I'm trying to show people my experience in life that would feel weird.}

\boldpar{C3: Preserving creative authorship}
Above all, creators emphasized maintaining creative vision and control ($n=5$)~\cite{polimetla2025paradigm,yeh2024ghostwriter}: \myquote{I want to have a say in the AI [and] use my own human creativity to get the story I'm intending to tell} (F5).
Three felt strongly about crafting story elements such as scripts themselves: \myquote{A very important thing to me is the voice behind the video... I put a lot of care into my script and making sure it's my tone and words} (F1). 
F7 added that writing \myquote{with ChatGPT [makes] it feel less my work,} especially for personal stories.

\figureScaffolded
\section{Generative Expansion: Context-Aware Story Expansion with Generative Video}\label{sec:design_goals}

Our formative study highlighted both the opportunities and challenges of using generative video to augment existing footage.
Building on these insights -- and an analysis of nine filmmaking texts that discuss practices from narrative theory and cinematography for expanding stories in coherent, engaging ways (see \autoref{app:books}) -- we propose \textbf{generative expansion}, a design framework for creating video stories through systematically blending captured and generated material (\autoref{fig:scaffolded}). 

The framework centers on context-aware generations that leverage both visual (\ie style) and narrative context to expand a video.
Generative expansion extends prior work centered on fully-captured \cite{marcelo2025editduet,huh2025videodiff} or fully-generated media~\cite{skyreels,xu2025filmagent}, advancing the underexplored space of hybrid video stories that blend both forms of media.
It comprises four interrelated design goals (\textbf{D1-4}) that navigate the tension between providing \textit{structured support} for narrative coherence and \textit{creative freedom} guided by user intent.

\subsection{\textit{Explore}: Surfacing Potential Story Ideas \textbf{(D1)}} 
Story construction typically begins with a premise or logline that crystallizes their central purpose~\cite{synder2005save,truby2007anatomy}.
Yet, our formative study revealed the difficulty of shaping coherent narratives from unstructured ideas and materials.
Systems should scaffold this process by surfacing narrative patterns such as themes~\cite{mckee1997story,truby2022anatomy,mckee2021character} or alternative arcs~\cite{lan2021understanding} latent in a user's footage, without prescribing a fixed storyline (\textbf{O1, O2}).

\subsection{\textit{Expand}: Highlighting Opportunities for Narrative Development (D2)}
Beyond structural ideation, systems should detect gaps where additional material could improve narrative coherence and flow (\textbf{O1}). 
Drawing on film grammar -- continuity editing~\cite{bowen2018grammar}, question–answer shot patterns~\cite{katz1991film}, rhythm and pacing~\cite{mckee2022action,mckee2016dialogue} -- systems can flag weak links and suggest connective shots for expansion. 
This reframes generative video from mere production to intentional, story-grounded augmentation. 

\subsection{\textit{Blend}: Generating Contextually Connected Video Clips (D3)} 
Once gaps are identified, systems should propose new clips to fill them, helping to bring unrecorded moments to life (\textbf{O2}). 
Suggestions should be grounded in cinematography principles that emphasize how shot types, angles, and framing convey distinct semantic meanings (\eg wide shots for context, close-ups for emotion)~\cite{bowen2018grammar,katz1991film}.
By adapting these practices, generated clips can meaningfully extend the story world and visually align with surrounding footage, rather than feel like arbitrary insertions (\textbf{C1, C2}).

\subsection{\textit{Control}: Steering Narrative Shape and Style Across Blended Media (D4)} 
While systems can suggest and supply new material, users should retain creative agency at each stage of storytelling (\textbf{C3}). 
Just as writing and cinematography rely on deliberate choices -- whether shaping theme and character~\cite{truby2007anatomy,mckee2021character} or framing a scene through shot design~\cite{bowen2018grammar} -- the system should provide intuitive ways to steer story sequence, shot composition, and video generation when blending captured and generative footage.
Generative expansion thus positions AI as a collaborator for story ideation and visualization while ensuring the hybrid video reflects the creator's narrative and stylistic intent (\textbf{O3, C2}).

\section{\system System}\label{sec:system_design}
Guided by our design goals, we developed \system, a video authoring tool that instantiates generative expansion for narrated video stories~\cite{truong2016quickcut,lu2023show}.
To study how different users engage with this emerging paradigm of hybrid video creation (\ie blending captured and generative footage), we designed our tool to support a broad range of creative control.
\system provides high-level AI scaffolds to organize footage and generate missing moments as well as fine-grained controls for refining model outputs for those who desire deeper involvement.

While our system includes basic editing operations like trimming or rearranging footage, its focus is on supporting story construction rather than full post-production. 
Accordingly, \system outputs a \textit{rough cut}~\cite{katz1991film,bowen2018grammar} -- an initial sequence of clips and audio that establishes narrative structure -- leaving final touches to professional tools such as Adobe Premiere Pro.

\figureInterface

\subsection{Interface}\label{sec:interface}
Our interface comprises two linked views that bridge the visual and voiceover components of narrated video stories. 
The \textbf{canvas} allows creators to (1) \textit{organize} and \textit{generate} visual materials (images or videos), and (2) \textit{shape} narratives by linking shot and scene nodes (\autoref{fig:interface}).
It transforms a traditionally freeform space into a semi-structured visual storyline, enabling direct visualization and manipulation of narrative flow like a video editing timeline.
The \textbf{script editor} is where users compose narration \textit{scripts} and story \textit{context}, which guide and personalize video creation (\autoref{fig:editor}).
Together, this dual-view design integrates the affordances of freeform ideation and linear editing into a unified workspace for story construction, supporting diverse video storytelling workflows (\autoref{sec:form_study_sub1}).

\figureEditor
 
\system also employs \textit{semantic zoom}, adjusting information density with scale to mirror the sematic levels of video creation -- story, scene, and shot~\cite{katz1991film,bowen2018grammar} -- and aid sensemaking~\cite{suh2024luminate,suh2023sensecape}. 
Zooming out collapses each scene into a keyframe (defaulting to the first shot) and reduces the editor to a scene outline (\autoref{fig:teaser}), supporting high-level story reasoning; zooming in stores scene-level detail and full script context.

\subsection{Feature Overview}\label{sec:walkthrough}
Next, we describe how \system's features operationalize generative expansion. 
Screenshots illustrate creating a video story about traveling to Japan for CHI 2025. 
For clarity, we introduce canvas and editor features separately, though in practice they interleave to support iterative creation (see user vignettes in \autoref{sec:rq1}). 
Since our goal is to scaffold rather than automate the creative process, all AI-powered features are suggestive, giving creators autonomy to accept, reject, or adapt outputs. 

\subsubsection{Getting Started}
Upon opening \system, creators can add notes to the editor to outline initial story ideas and upload captured media to the canvas as shot nodes (\autoref{fig:group}a). 
As nodes are added, \system analyzes their content to produce rich text descriptions that later inform narrative suggestions (\autoref{fig:interface}c).

\subsubsection{Canvas}
In the canvas, creators craft their visual storyline.

\boldpar{Creating a story sequence}
To help, \system's \textbf{group} feature automatically organizes shots into scenes, assigning each a semantic title, color, and description based on the user's notes (\autoref{fig:group}b). 
Scenes can then be adjusted (\eg renaming or moving visuals) and \textbf{sequenced} into a possible storyline (\autoref{fig:group}c, \explore). 
When sequencing, \system visualizes narrative gaps by proposing new scenes to connect existing moments, such as ``Anticipation for CHI Japan'' and ``Conference Connections and Breaks'' (\expand).
Creators can also add \textbf{contextual transition scenes} to expand the narrative.

In \autoref{fig:new_scene}, pressing the \textit{plus} button between ``Yokohama Culinary Delights'' and ``Kamakura Town Charms'' yields ``Exploring Yokohama's Landmarks.''

\boldpar{Comparing story variations}
To support exploration, \system enables \textbf{story variations}, where users can prompt for alternate storylines (\eg ``focus on nature imagery''). 
With \textit{compare mode}, creators can visualize story versions side by side, and receive a tailored analysis of their strengths and weaknesses (\autoref{fig:compare}, \explore). 

\boldpar{Expanding scenes}
Next, creators can connect shots within each scene to storyboard visuals. 
To assist, the \textbf{sequence visuals} feature (\autoref{fig:visual_sequence}a) automatically proposes an engaging sequence from existing shots and generates new ones to (1) fill gaps and (2) add different perspectives through varied shot types and angles (\autoref{fig:visual_sequence}b, \expand). 
Shot suggestions are outlined in \textcolor{AIColor}{orange} (vs. \textcolor{RealColor}{purple} for uploaded shots) and include explanations of how they enhance the narrative.
Connections between visuals can be enhanced with \textbf{contextual transition shots} (\autoref{fig:visual_sequence}c, \blend).

\figureGroup

\boldpar{Customizing shots}
Any static image can be turned into video with the pink \textbf{animate} button (\autoref{fig:interface}).
Users can further tailor outputs to their creative vision in the \textbf{video generation} panel (\autoref{fig:controls}b).
Here, they can add \textit{spatial annotations}~\cite{spatialprompt} to illustrate desired story elements (\eg ``add birds flying over Buddha's head''), similar to how professional filmmakers mark up frames~\cite{katz1991film}, and press ``suggest'' to auto-populate \textit{structured prompt fields} covering common cinematic dimensions (camera movement, lighting, style, action)~\cite{bowen2018grammar} based on the image and annotations (\control). 
Images can also be refined in the \textbf{image editor} through prompting (\eg ``Transform the sky into sunset'') before they are brought to life (\autoref{fig:controls}a). 

\figureVisualSequence

\subsubsection{Script Editor}
Complementing visual story construction, the script editor scaffolds narration writing.
Creators can \textit{sync} their original story notes to the script (\autoref{fig:editor}e), which automatically segments them by scene to provide a starting point. 

\boldpar{Expanding the script}
Rather than drafting a full script, \system's \textbf{suggestions sidebar} provides Socratic-style questions and prompts~\cite{gmeiner2025exploring,elder1998role} to help creators expand their own ideas (\autoref{fig:editor}a, \explore, \expand).
For example, ``How did the live calligraphy demonstration make you feel, and what specific details captured your attention?'' (\autoref{fig:editor}b).
Users can generate broader story-level and specific scene-level suggestions, which span narrative categories such as imagery, structure, pacing, and theme (\autoref{tab:categories}).
Each suggestion comes with an explanation and tips.
Pressing ``address'' on a tip generates three refined versions in the \textit{inline refinement menu}, which users can build on to update the script.
Creators can help the system learn their preferences by disliking suggestions~\cite{yeh2024ghostwriter} (\control).

\boldpar{Adding voiceover and music}
Users can add voiceover audio and background music to each scene to enhance the story (\autoref{fig:editor}).

\subsubsection{Connecting Visuals and the Script}
To integrate the visuals and script for each scene, \system provides a \textbf{scene timeline} (\autoref{fig:timeline}), which displays both in a timeline editor mirroring tools such as Premiere Pro and DaVinci Resolve. 
Users can rearrange footage, adjust segment widths, and use the ``auto align'' feature to explore different ways of synchronizing visuals with the narration (\control). 
The evolving rough cut can be viewed at any time via the preview video button (\autoref{fig:interface}).

\subsection{Context-Aware Video Generation}
A core capability of \system is generating contextual video clips that blend with captured footage narratively and stylistically.
To do this, we implemented a two-stage pipeline:
\begin{enumerate}
    \item \textbf{Contextual keyframe generation:} \system first generates a starting keyframe for the new shot, conditioned on \textit{surrounding shots} (previous and next) for visual continuity, as well as the \textit{story context} and \textit{scene script} for story coherence (\autoref{fig:generative_pipeline}a).
    Users can optionally provide a \textit{prompt} to guide creative intent.
    As intermediate output, \system produces three \textit{image generation prompts}, which ideate different shots to expand the story.
    Variations are informed by the cinematography principles we derived for building a compelling visual narrative (\eg shot composition, lighting, framing -- see \autoref{app:books}).
    Finally, \textit{three keyframe candidates} are generated from the image prompts, which the user can adjust and select from.
    \item \textbf{Video animation from keyframe:} Once a keyframe is selected, \system animates it into a video clip (\autoref{fig:generative_pipeline}b).
    Users can provide a \textit{prompt} or \textit{annotate} the keyframe directly to guide how the motion and scene should unfold.
    Drawing from the same cinematography principles, \system produces an intermediate \textit{augmented video generation prompt} based on these inputs.
    This prompt is used to generate $N$ (default: 2) video variants.
    The original (unannotated) keyframe is used for generation to avoid visual artifacts.
\end{enumerate}

\figureTimeline

\noindent Additional feature and prompt details for \system are included in \autoref{app:pipeline} and Supplementary Materials. 
Each AI call is conditioned on the user's evolving visual storyline and script -- story context, scene descriptions, file metadata -- and combined with our narrative and cinematic principles to ground system suggestions in filmmaking theory.

\figureGenerativePipeline

\subsection{System Implementation}\label{sec:implementation}
\system is a full-stack application with a React/TypeScript frontend and Python/Flask backend. 
The canvas interface is built with React Flow~\cite{reactflow}, and the script editor with Tiptap~\cite{tiptap}.
LLM features are powered by Google Gemini (\texttt{gemini-2.5-flash})~\cite{geminiflash}, including scene creation and sequencing, story variation, script suggestions, visual sequencing, and timeline auto-alignment.
We use Gemini's image model (\texttt{gemini-2.5-flash-\allowbreak image}, code name \textit{*Nano Banana*})~\cite{geminiimage} to generate images, and Veo3 (\texttt{veo-3.0-fast-\allowbreak generate-\allowbreak preview})~\cite{veo3} for videos.
Music and speech are generated by custom models trained on licensed instrumental-only music and speech. 
Image generation takes $\sim$10s and video generation $\sim$40s, with calls parallelized to reduce latency.

\section{User Study}\label{sec:user_study}
To examine how \system supports video storytelling, we conducted an exploratory study with 12 creators, each producing a short narrated video using our system, following methods similar to prior evaluations of creative authoring tools~\cite{cao2025compositional,cao2023dataparticles,jahanlou2022katika}. 
Our goal was to understand how different users perceive and interact with \system and its ability to blend captured and generative video, rather than to evaluate final output quality.
We did not include a baseline because although many generative video tools are emerging, their capabilities and workflows vary widely, and we were not aware of any single tool that supports hybrid workflows comparable to our generative expansion approach. 
Our study was guided by three research questions:

\begin{enumerate}[start=1,label={\bfseries RQ\arabic*.}] 
    \item How do creators approach and adapt their video storytelling workflows when using \system? 
    \item How do creators perceive \system's features as supporting (or challenging) the development and expansion of video stories compared to existing practices? 
    \item How does blending captured and generated media impact the storytelling process? 
\end{enumerate}

\subsection{Experimental Setup}
\boldpar{Participants}
We recruited 12 video creators \textbf{(P1–P12)} through internal mailing lists and snowball sampling, seeking diversity in experience and genre. 
Participants included eight employees from \location, three PhD or university students, and one social media manager (\autoref{tab:study_participants}).
Three self-identified as professional video editors (\textit{avg:} 14.7 years of experience), five as intermediate (\textit{avg:} 5.8 years), and four as novices (\textit{avg:} 3.4 years).  
Eight had prior experience with generative video tools such as Veo3 or Firefly (\textit{avg:} 1.1 years), while four were new to them. 
Creators produced a wide variety of videos during the study, including travel vlogs, wedding reels, and personal essays.

\boldpar{Task}
Each participant created a rough cut of a $\sim$1–2 minute narrated video using \system, working with their own materials to keep the task personally meaningful and aligned with their typical creative practice.
Creators submitted their media and story idea in advance, which was preloaded into the system.  
We did not restrict the number or type of content to better assess \system's performance across varied inputs, but suggested $\sim$40 files given the study's duration. 
In practice, participants brought 10-77 images and/or videos.

\subsection{Study Procedure}
Sessions were conducted over video call and lasted 1.5 hours. 
We asked creators to think aloud, collected screen and audio recordings (with consent), and logged system events. 
Participants received \$60 in compensation. 

\tableStudyPart
\boldpar{Pre-task survey \& interview ($\sim$10 min)}
Before the study, participants completed a brief survey about their background with video making and generative video tools (reported above). 
At the start of each session, they elaborated on their answers and shared initial perceptions of incorporating generative footage into their workflows. 
Nine creators expressed excitement, while three were more hesitant, citing concerns such as style inconsistencies or negative audience reactions.
Four anticipated using generative media to fill gaps and extend content, and three highlighted its potential to surface new story ideas or perspectives from their footage.

\boldpar{Structured walkthrough ($\sim$30 min)}
We introduced \system with a structured walkthrough.
Due to its relative complexity, we wanted to ensure participants had ample time to familiarize themselves with our system.
Creators watched a three-part tutorial video (\textit{total}: 10 min), where each part covered one set of features. 
After each segment, participants practiced using those features on their own materials for $\sim$7 minutes:
\begin{enumerate}[start=1,label={\textbf{\Roman*.}}] 
    \item \textbf{Ideation and Exploration.} First, participants learned how to arrange assets, create and sequence scenes, and generate story variations (\explore).
    \item \textbf{Narrative and Visual Expansion.} Next, creators learned features related to \expand~ and \blend, such as expanding the script with story suggestions, or expanding the visual sequence of a scene with new contextual shots.
    \item \textbf{Refining and Aligning Outputs.} Lastly, we introduced our creative controls for refining image \& video generations, along with the scene timeline for aligning visuals to the script (\control).
\end{enumerate}
The order of features was fixed to reflect a natural learning sequence and reduce cognitive load for first-time users.

\boldpar{Freeform exploration ($\sim$30 min)}
Participants then freely developed their stories using \system's features.

\boldpar{Post-task survey \& interview ($\sim$20 min)}
Finally, creators completed a Likert-scale survey assessing feature utility across tasks and phases of video storytelling, as well as overall satisfaction with \system. 
Items were adapted from the System Usability Scale (SUS)~\cite{brooke1996sus} and Creativity Support Index (CSI)~\cite{cherry2014quantifying}. 
We also conducted semi-structured interviews to gather reflections on how \system compared to existing workflows, perceptions of blending captured and generated content, and suggestions for improvement.

\subsection{Data Analysis}
We adopted a mixed-methods approach for data analysis. 
From screen recordings and system logs, we collected a quantitative account of feature usage (\eg number of images/videos generated, story suggestions clicked). 
Qualitatively, we analyzed responses to post-task Likert-scale questions and synthesized themes from think-aloud and interview transcripts.

\section{Results}\label{sec:results}
Overall, creators enjoyed developing video stories with \system and expressed interest in using it again (\autoref{fig:likert}).
While some noted the system was not the easiest to use ($n=5$), describing it as \myquote{challenging but learnable} (P2), participants valued \system's flexibility -- allowing them to balance generative suggestions with their own ideas (P7) -- and described the experience as \myquote{really cool [and] fun} (P9) and even {addictive} (P5, P12).
Eleven emphasized that \system \textbf{augmented their creativity} during storytelling: \myquote{I like how it leveraged my ideas and expanded upon them [but] I was still holding the steering wheel} (P6).
P10 even asked, \myquote{can I share the video I made on social media?}

\figureLikert
\subsection{RQ1: How do creators approach and adapt their video storytelling workflows when using \system?}\label{sec:rq1}
We illustrate emerging workflows through two representative vignettes, showcasing how a novice (P2) and professional (P4) video creator integrated generative expansion into their creative process.

\subsubsection{P2: Reliving Uncaptured Moments and Exploring Substories}\label{sec:casestudy_one}
P2, an amateur creator, wanted to make a video story about her 2025 summer highlights, including a surprise engagement and Ireland trip. 
She first grouped her media into scenes and asked \system to automatically sequence them (\explore). 
To her surprise, \system's suggestions closely matched her envisioned storyline while omitting extraneous scenes: \myquote{The home projects and dance recital was... a summer highlight, but it was kind of thrown in there. It's cool that it's like if we're telling a strong story, maybe that doesn't fit and addressed that.}

P2 then viewed the story suggestions (\eg ``How did the proposal feel?'' or ``What was an unexpected joy or growth throughout the summer?''), finding them \myquote{solid recommendations for storytelling, [which is] specifically something I need help with} (\expand). 
She augmented her script with these suggestions, starting from the ``Surprise Engagement \& Celebration'' scene (\autoref{fig:p2}a).

Next, she built out the scene using the sequence visuals feature, where \system added three new shots to expand the initial two videos (\blend): \myquote{It's cool to see how it's taken the little I provided [and] created more of a story. That's crazy. 'Cause that's what my parents were wearing that day... That's definitely something I would use.} 
She reflected, \myquote{it's mind boggling that it's even a possibility to have content when we were so present that we didn't have our phones. It's AI generated, but it's pretty representative... Like we did all cheers [and] sit at the table with family around.} 
Finally, using the scene timeline, P2 aligned the visuals to her script and recorded narration (\control).  

\figureCaseStudyOne

Later, P2 created a story variation focused only on her Ireland trip (\explore).  
Here, \system identified the relevant media and broke it into multiple granular scenes, whereas the original sequence grouped her Ireland highlights into two larger scenes (\autoref{fig:p2}b). 
\system also suggested contextual scenes to fill narrative gaps, such as ``Irish Trip Departure'' and ``Reflections on a Summer Chapter.''
P2 shared, \myquote{That's pretty cool that it connected the dots and drew a story out of just the Ireland footage,} and continued expanding this new storyline. 

\subsubsection{P4: Finding New Inspiration Through Generated Shots}\label{sec:casestudy_two}
P4, a professional videographer, brought raw footage from his friend's wedding to explore. 
He began by having \system automatically group and sequence his clips (\explore), making adjustments to align with his creative vision (\eg removing an extraneous ``Elegant Cocktail Hour'' scene). 
After generating story suggestions (\expand), P4 found the \textit{structure} prompt -- ``Consider starting the story with [the bride]'s anxiety about the horseback entrance to immediately introduce conflict'' -- intriguing, but worried whether \myquote{I have enough footage to build this. Cause here I only have two videos.}

However, feeling inspired, P4 used sequence visuals to expand the ``Anxious Arrival \& Horseback Entrance'' scene (\blend). 
\system added a wide shot of the couple entering on horseback and a medium shot of the bride (\autoref{fig:p4}a). 
Using annotations, P4 animated the medium shot, sketching the groom's entrance and requesting a video prompt suggestion that included ``a gentle breeze rustling leaves'' and ``white butterflies flitting past'' (\control). 
He found the generated shots \myquote{really nice}, but \myquote{I don't think I would use them, especially for wedding videos where the point is to capture real memories.} 
P4 also observed that to mimic his own \myquote{videos [which] are in log,} the AI generated \myquote{washed out footage, but these aren't actually log [so] I couldn't color grade them [the] same way.}

Next, P4 explored the ``Sacred Church Ceremony'' scene, where he ran sequence visuals again, observing:
\begin{quote}
"[\systemx] puts the settings at the beginning like two establishing shots. And the generated image here is the hand of probably the bride. It creates interesting tension [through] only showing parts first and maybe revealing later. That's cool to build suspense... I could do slow pushing and the hands look nervous and touch. The idea could be the bride is stressed again and we're only seeing her hands..."
\end{quote}
which he acted on to animate the image into a video (\autoref{fig:p4}b).
Further examining the suggested sequence, P4 found additional inspiration for expansion:
\begin{quote}
"After this, you know what would be cool? I want [to] add a wide shot with the two characters in the middle, not highlighted, but making sure they're the emphasis... So starting from settings to little parts of the scene to the full scene to create a real shock or boom, this is where we are."
\end{quote}

\figureCaseStudyTwo

\subsection{RQ2: How do creators perceive \system's features as supporting (or challenging) the development and expansion of video stories compared to existing practices?}\label{sec:rq2}
Creators found various features helpful at different phases of story development (\autoref{fig:likert}), especially as compared to existing workflows.
In particular, \system was perceived as useful for narrative exploration and expansion, reflecting favorably on our design goals, with novices leaning more on suggestions, while experts selectively used or refined them.

\subsubsection{Strength: A Unified Workspace for Video Storytelling}
A recurring theme was \system's value in offering a single, unified workspace for video creation ($n=5$), which echoes our formative study findings (\autoref{sec:form_study_sub1}). 
P6 noted, \myquote{Everything is already integrated [with] the script and storyboard. That's huge. I can't do that normally.} 
Similarly, P7 described \system as a clear improvement over her current workflow: \myquote{Normally, I need like 5 AIs... here you can do it in one place and combine everything [to] create a full story.}
Participants also praised the synchronization of \system: \myquote{I love the notes and being able to work and structure it there, but then have it reflect visually [in the canvas]} (P12). 
Three highlighted the ability to automatically align visuals with the script as a powerful augmentation of existing tools (\control): \myquote{The timeline thing in its own is crazy. I don't know why it's not in Premiere yet} (P8).

\subsubsection{Strength: Story Visualization and Manipulation in the Canvas}
Both novice and experienced creators valued how \system organized and visualized their assets into cohesive stories ($n=8$).
Four highlighted the novelty of our \textbf{semi-structured canvas} for video creation, drawing parallels to Figma (P8) and Miro (P9).
P2 reflected, \myquote{Video editing has been so limited to the timeline... it's exciting to move videos around and see how they can be sequenced in different ways,} emphasizing how \system supports new forms of story visualization and manipulation.

In particular, P11 called the contextual \textbf{scene grouping and sequencing} features \myquote{game changers} and P1, who made a video about visiting the Nintendo Store, emphasized that \myquote{having it automatically say, here's the exterior, here's the Zelda stuff, here's the Pokémon. That's gold... I usually have to do that manually.}
Others appreciated how automatic grouping reduced creative friction: \myquote{I get overwhelmed by my files and it impedes me from even wanting to create. So having that effort taken out of it... I loved that} (P12).
P4 added, \myquote{Seeing how AI grouped my content helped [paint] a clear visual of what stories I could tell and made potential narratives I didn't consider more salient.}

\subsubsection{Strength: Visual Story Expansion}
Ten creators found \system's visual suggestions helpful for story expansion (\autoref{fig:likert}). 
They generated an average of 31 images (\textit{min:} 4, \textit{max:} 65) and 8 videos (\textit{min:} 1, \textit{max:} 17) per study session, augmenting their original media by $\sim$96.4\%. 

Novices appreciated how \system filled narrative gaps with \textbf{context-aware shots} that felt creative and customized (\blend).
P9 loved \myquote{how accurate they are to the story and how they add more to it. I wouldn't have thought to add the two turtles in the shadow of the seagull... and [here] it zoomed in on the same seagull's foot.} (\autoref{fig:storyboard}b).
P6 added, \myquote{It's scary accurate. What the heck? This is literally my desk and I have the same plant.}

More experienced creators leveraged \system's sequencing and generation suggestions to test ideas and spark new directions.
P3 was impressed when two existing visuals were aligned into a match cut, reflecting our distilled cinematic frameworks~\cite{bowen2018grammar}: \myquote{It fits nicely with this segment [to] show the passage of time} (\autoref{fig:storyboard}a).
Creators found \system's suggestions especially \myquote{helpful for storyboarding ideas quickly} (P5) and \myquote{generating what comes in between two shots... Being able to scaffold what the scene might look like and situate that progress in time is neat} (P10), again speaking to our canvas' role as a visual storyline (\expand).

\subsubsection{Strength/Limitation: Narrative Expansion Through Script Suggestions}
Script suggestions were generally perceived as helpful, but some creators had mixed feelings, particularly during narrative ideation (\autoref{fig:likert}).
Each participant clicked an average of 3.7 suggestions (\textit{min:} 1, \textit{max:} 11) and used ``address'' an average of 2.0 times (\textit{min:} 0, \textit{max:} 6).

Suggestions -- both story and scene -- were typically more helpful for novices or those with less concrete ideas (\expand).
P9 thought they provided \myquote{nice starting points [for] thinking. I struggle with writing so the suggestions [helped] elevate my storytelling.}
P8 said, \myquote{I don't like AI generated scripts, but having someone push you to think about character, imagery, emotion, that was cool.}
Conversely, creators with a stronger vision (often experts) felt some narrative suggestions did not comprehend or challenged their ideas: \myquote{I like having the autonomy [to say] this is what I want to create, rather than it saying, oh, right now your story kind of sucks} (P3).
P4 noted that the utility of a suggestion \myquote{really depends... I want [to give] my motivations or discuss it like when we brainstorm with people.} 

\subsubsection{Limitation: Refining Generative Outputs}
While creators enjoyed generating new media with \system~ and found our refinement controls valuable -- with three praising the annotation feature as \myquote{especially cool... I've never seen anything like that} (P12) -- six still found it difficult to \textbf{refine outputs} to match their intent (\autoref{fig:likert}). 
Refinement was especially challenging for novices, who did not iterate or engage as much with these features, emphasizing the persistent struggle of steering generative models (\control).
P2 said \myquote{the hardest part was getting the look and feel [of] generations right,} and P11 felt that our system \myquote{needs a lot of prompts and I kind of don't like that.}

\subsection{RQ3: How does blending captured and generated media impact the storytelling process?}\label{sec:rq3}
Creators were largely excited about augmenting footage with generative video ($n=9$): \myquote{Even having it as an option opens up so many new storytelling possibilities} (P9).
P12 said he \myquote{would literally use the [blending] aspect in every scene, maybe every shot.}
However, four felt \myquote{the main beats of the story need to be my own footage} so any generated shots \myquote{should be supplemental} (P3).

Below, we share use cases for blending captured and generative media, limitations of this approach, and its influence on perceived creative ownership, as surfaced through \system.

\subsubsection{Use Cases for Augmenting Existing Media with Generative Content}\label{sec:use_cases}
Overall, participants valued how \system generated new media \textit{grounded} in their existing story and footage, and \textit{visually sequenced} them in ways that sparked storytelling inspiration beyond captured materials.

Three expressed how \system's suggestions helped them \textbf{translate abstract reflections into a concrete visual form}: \myquote{I was pleasantly surprised by the binoculars and person pointing. That's not necessarily an image I had in my head, but it's nice because these were vague memories... and the generations allowed me to put [that] on paper} (P3, \autoref{fig:storyboard}a).
Similarly, P10 described how \system helped bring her memories to life through augmenting captured moments: \myquote{I love this. I don't see my chef friends often anymore so I was like how do I pretend I'm in the kitchen again and it just auto-generated this image that allows me to do that} (\autoref{fig:storyboard}c).

Six leveraged \system's creative controls to \textbf{add visual depth and cinematic layers} to their story.
For example, P4 and P12 used annotation prompts to insert new story elements -- such as the groom walking into a shot (\autoref{fig:p4}a) or a deer entering a forest (\autoref{fig:creative_storytelling}a) -- into existing frames.

\system's visual suggestions also encouraged creators to \textbf{explore different perspectives and narrative directions} ($n=5$): \myquote{Being able to see what it comes up with is really helpful. That would change the way I create} (P1).
P9 loved how the system \myquote{kept giving me new ideas... like the shadow of the seagull, the emotion on the turtle's face} contextualized to her story.
Watching \system sequence his existing visuals and suggest new shots, P12 was inspired to add an aerial drone shot, which would have been otherwise impossible to capture: \myquote{Obviously I'm not gonna get that shot normally cuz what am I gonna do? Throw my iPhone in the air? So the fact that generative video could add that was really cool} (\autoref{fig:creative_storytelling}a).

\subsubsection{Limitations of Blending Media for Video Storytelling}\label{sec:limitations_blending}
Despite excitement about \system's blending capabilities, creators also expressed concerns about incorporating generative visuals into their videos.

\boldpar{Challenges with blending captured and generative visuals}
Two worried about social stigma around AI-generated media: \myquote{A lot of creatives are threatened by AI... even if you get a sense of someone's using AI, you get canceled low-key} (P8).
Four, who made videos about real experiences, were concerned about \textbf{portraying fabricated moments}.
P3 shared, \myquote{It doesn't feel real if I'm generating a fake image [of] a meal. These are fake memories of people eating... Even if it fools someone else, it wouldn't feel authentic.}
P1 explained that acceptability depends on the video content and goal.
For his Nintendo video, \myquote{I'm expecting specifically to see my kids, Mario, Donkey Kong and Pikachu. So when it doesn't get it 100\% right, it feels dishonest. If it was an establishing shot of San Francisco or trees in a forest... then there's a little more leeway.}

Three expert creators reflected how blending media to fill narrative gaps could \textbf{potentially dilute creativity} as \myquote{the beauty of creating a video comes from the fact that you can't have everything... you piece things together and with editing and music, you're surprised by what emotions you can create} (P3).
P12 added, \myquote{Often a good story doesn't give too much -- that's why people say show, don't tell.}

\boldpar{Challenges from generative model capabilities.}
P8 described how \myquote{sometimes the generations just looked kind of weird and [didn't] match the story,} highlighting the larger challenge of \textbf{vision and ``vibe'' misalignment.} 
During P1's study, \system generated a comedic clip of Bulbasaur blinking, which he described as \myquote{a fun serendipitous thing. But I obviously can't use this... if I did it would be in a joking way}. 
Similarly, P4 noted that \myquote{the [video prompt] suggestions are very dramatic. It turns everything into Hollywood VFX [like] a flurry of shimmering white butterflies, dogs and fireworks.}
In some cases, style shifts were desired, like in P5's fantasy story (\autoref{fig:creative_storytelling}b), 
but for more realistic videos \myquote{if [the generative content] takes the viewer out of the video, that's a problem} (P12).

P4 also reflected how although stylistically, the generations could be convincing, other aspects limit their practical use.
Beyond the log footage issue (\autoref{sec:casestudy_two}), P4 found in some generated clips, \myquote{the camera movement is not something I was able to have cause I didn't use a gimbal.}
He noted other inconsistencies such as how he \myquote{didn't get to shoot at that time [suggested by the system]} or \myquote{have such a long focal length} during the wedding. 

\subsubsection{Impact of Blended Media on Creative Ownership During Storytelling}\label{sec:ownership_results}

Eight participants felt adding generative content \textbf{would not impact} their perceived ownership over the final video due to the \myquote{freeform and flexible} nature of \system (P8).
P5 explained, \myquote{AI doesn't reduce ownership. I'm always going to do extra touches. [It's] just another tool in the toolbox... I'm not expecting nor do I want it to take over.}
P6 also felt he retained creative control: \myquote{It gave me suggestions, a starting point. But I was saying no, I don't agree with this... not everyone can use AI effectively and communicate with it. It's all about that lens, perspective, and taste.} 

Three said they \textbf{would feel less} ownership.
P9 explained, \myquote{I don't really consider me prompting something as my own... I wouldn't feel authentic creating generative content and claiming it as mine.}
P3 added that \myquote{particularly for personal videos, I worry about maintaining autonomy and making it still feel like my story.} 
P4 said his perspective depends: \myquote{Generating an image of a hand isn't the core element of the story [but] if the full narrative was generated by AI, I wouldn't feel ownership.}
He added, \myquote{For some AI-generated videos, I [spend] 100 hours on a single clip. It's hard not to feel ownership... but if everything is being generated in 30 seconds, I'm not gonna feel very attached.}

\section{Discussion}
Based on our study findings, we reflect on key opportunities for blending captured and generative video, current challenges, and our vision for future video storytelling interfaces.

\subsection{Expanding Stories Through Blending Captured and Generative Video}\label{sec:comparing_workflows}
Our results highlight distinct affordances from \textit{blending} captured and generative media. 
Whether filling narrative gaps with uncaptured moments or visualizing otherwise impossible shots, generative expansion enabled creators to extend stories in creative ways while staying \textbf{grounded and coherent} within their existing materials (\autoref{sec:use_cases}).
Notably, incorporating generative clips did not diminish most creators' sense of ownership (\autoref{sec:ownership_results}).

\subsubsection{Beyond Temporal Blending}
Our work foregrounds \textit{temporal} blending -- bridging gaps across shots in time for narrative coherence -- as one point in the broader design space of integrating captured and generative video (\autoref{fig:design_space}).
Emerging generative techniques suggest opportunities for \textit{spatial} blending within a single frame (\eg filling gaps in environments behind real actors with generative VFX~\cite{liu2025vfx}).
We also see potential in \textit{stylistic} blending, where the primary goal is expressivity over continuity~\cite{liu2024personalised}.
As illustrated by P5's fantasy video story (\autoref{sec:limitations_blending}), mixing styles across time (between shots) and space (within a shot) opens further storytelling possibilities. 

\subsubsection{Comparing Captured, Generative, and Blended Workflows}
Blended workflows promise to unite the authenticity of captured footage with the flexibility of generative media (\autoref{fig:design_space}).
Yet creators often treated generated clips as \textbf{second-class or supplemental} to captured material (\autoref{sec:use_cases}).
This highlights an interesting design tension: generative content expands what is possible, but is perceived as illustrative rather than authoritative in shaping video narratives.
Such perceptions may be tied to how users understand authorship in AI-infused video creation.
Future work could examine how to elevate the role of generative video in meaningful ways, while preserving creative ownership and integrity of captured material.

Blending also accentuated imperfections of generative footage -- \eg style or color mismatches with captured media (\autoref{sec:limitations_blending}).
While similar challenges arise in fully captured or generated workflows~\cite{xiao2025videoauteur,qu2024exploring,skillman2025pros,allinmotion2023stock}, they are especially salient in blended settings, as outputs sit side by side~\cite{wang2025boundary}.
These issues may fade as models improve, but they underscore a key consideration for minimizing disruptions to a creator's vision -- and their story.
A systematic comparison of captured, generative, and blended workflows could clarify how tensions surface across different storytelling contexts and user segments.
Finally, while we focused on augmenting captured footage, it would be valuable to explore how generative expansion extends to other forms of initial media, including generated or rendered video.

\figureDesignSpace

\subsection{Shaping Generative Creativity Through Constraints}
Our goal was to help bridge narrative gaps and expand video storytelling possibilities.
However, as some participants noted, the unlimited creativity promised by generative models may not always be desirable since \textbf{creativity often comes from having constraints}~\cite{rosso2014creativity,acar2019creativity}.
Thus, we see opportunities to refine our generative expansion framework.

\subsubsection{Adapting to the Creator's Reality}
Beyond stylistic inconsistencies, generated clips sometimes did not fit the ``vibe'' or filming setup of existing footage, constraining their practical use, particularly for professional creators (\autoref{sec:limitations_blending}).
These challenges highlight the need for generative models to adapt more closely to the user's footage and reality, possibly through additional contextual grounding, or personalized training/fine-tuning~\cite{wang2024animatelcm,liu2024personalised,xu2025personalized}.
Current systems often optimize for cinematic spectacle~\cite{veo3,sora}, but creators may prefer understated, imperfect clips to match what was feasible to capture.
This underscores another asymmetry between written and video storytelling: while text can often be rephrased without changing intent, subtle edits to a video can fundamentally alter meaning and tone.

\subsubsection{Learning the Poetry of Gaps and Missing Moments}
\system is designed to fill missing moments, but leaving gaps in a story can also strengthen its resonance and sense of poetry~\cite{mckee1997story,mckee2016dialogue}.
Next-gen AI video authoring systems should carefully consider which gaps to bridge and which to preserve.
In a future where creator-model alignment is solved, blending media may become a more intentional practice of weaving captured authenticity with generative possibility.
As users expressed discomfort sharing videos with fabricated additions (\autoref{sec:limitations_blending}), systems could deliberately withhold generations, learning from creators' narrative styles and allowing them to leave space -- both visually and verbally.
Doing so may enhance video pacing~\cite{wang2023improving}, preserve narrative integrity, and reduce memory distortion~\cite{pataranutaporn2025synthetic}.

\subsection{Future Interfaces for AI-Augmented Video Storytelling}
While our findings are limited by study scope, they highlight promising design directions for AI-assisted video storytelling interfaces.
Future work can also extend \system to explore systems tailored for specific expertise levels or content domains that blend captured and generative content.

\subsubsection{The Canvas as a Video Studio}
Creators valued our novel semi-structured canvas that unified script, assets, and generative models for video authoring (\autoref{sec:rq2}).
Many emerging tools offer infinite canvases for generative creation~\cite{superstudio,flora,midjourneypatchwork,fireflyboards}, but few, if any, include a direct video editing layer.
This suggests an avenue for future work: designing a \textbf{fully unified, multimodal canvas} for video creation.
What would it look like if all workflow components -- from asset generation and sequencing to script writing and clip editing -- coexisted in a single environment?
Malleable UI components~\cite{min2025malleable,cao2025generative} could also enable creators to reshape the interface around their own workflow, transforming canvases from auxiliary ideation spaces to central, adaptive video studios.

\subsubsection{Automation vs. Scaffolding}
Another core design question is \textbf{how much agency} systems should assume during video storytelling. 
Following generative expansion, \system offers narrative suggestions, but creators retain authorial control.
In contrast, fully automated tools generate complete scripts and visuals on the user's behalf~\cite{write-a-video,xia2020crosscast,cao2025generative}. 
Each approach carries tradeoffs: automation can accelerate editing and reduce barriers~\cite{truong2016quickcut,zhang2016snapvideo,2017leakeComp}, but risks generic outputs and diminished ownership~\cite{yeh2024ghostwriter,yuan2022wordcraft}.
Scaffolding preserves human creativity and autonomy, while demanding more effort~\cite{tilekbay2024expressedit,dhillon2024shaping,soe2021ai}. 
Creators also experience these tradeoffs differently.
As we saw, novices may prefer automation over fine-grained control, while experienced storytellers may favor light scaffolding that protects their narrative agency.
Preferences can shift during the creative process as well.
Future video creation systems could sense when creators want automation or scaffolding and fluidly transition between modes~\cite{lin2023beyond}.

\subsubsection{Feedback as Conversation}
As one form of scaffolding, \system includes Socratic suggestions to help expand video scripts.
However, creators with strong-formed visions found them less useful (\autoref{sec:rq2}).
This tension reflects how creativity resists prescription, and effective storytelling comes from knowing when to bend or break the rules~\cite{truby2022anatomy}.
We also observed experienced creators wanting to discuss \system's ideas, underscoring the value of \textbf{feedback as dialogue} over static suggestions.
While we chose not to implement a conversational agent to prioritize writing flow and creative authorship (\autoref{sec:form_study}), 
prior work shows how interactive feedback can support writing~\cite{lin2020enhancing,guo2022using}, design~\cite{zhou2024understanding}, and video understanding~\cite{zhang2025facilitating}.
Extending \system toward dialogue could further personalize support and enrich reflection, echoing the broader shift toward conversational creativity support tools~\cite{rezwana2025human,lin2023beyond}.

\subsection{Reflection Through Blended Storytelling}
Building on the discussion above, we believe blending captured and generative media (\autoref{sec:comparing_workflows}) invites new forms of reflection during video storytelling -- aligning with work on AI-augmented journaling~\cite{angenius2022talkus,kim2024mindfuldiary, pataranutaporn2024future,zulfikar2025resonance} and meaning-making~\cite{wang2025designing,schulz2024impact,torres2024design}.
Research shows that generating personalized text~\cite{kim2024diarymate}, music~\cite{park2025nore}, or images~\cite{park2025reimagining} can deepen the process of reflection.
By transforming photos into stylized AI images,~\citet{yun2025exploring} found that users re-experience and articulate emotions more vividly than with raw photos.
Although reflection was not our primary focus, we observed early signs of similar effects when creators relived past experiences through generative video (\eg P2's engagement story, \autoref{sec:casestudy_one}).
Future work could examine how blended storytelling can foster reflection while mitigating risks such as memory distortion and false recollection~\cite{pataranutaporn2025synthetic}.

\section{Conclusion}
We present \system, a video authoring tool that augments captured media with generative video to create hybrid video stories through the framework of generative expansion.
Grounded in narrative and cinematic principles, \system offers suggestions for expanding a user's story and generates contextually connected clips to bridge narrative gaps.
In a study with 12 video creators, we found that \system enables new expressive possibilities, such as bringing images to life in creative ways and crafting otherwise impossible shots, through systematically blending captured and generative content.
Future directions include exploring additional opportunities for blended storytelling and reimagining video authoring through adaptive canvas interfaces.
We hope our work inspires new avenues for AI-augmented video creation, allowing generative video to amplify human creativity in meaningful, narrative-driven ways.

\begin{acks}
The authors thank our fellow STORIE lab members and interns at Adobe for their invaluable support and feedback throughout this work. 
A special thanks to Gabi Duncombe for early input on our system design.
Additionally, we would like to thank the anonymous reviewers for helping shape this work, along with those who took the time to participate in our formative interviews and user studies.

\vspace{10pt}
\end{acks}

\bibliographystyle{ACM-Reference-Format}
\bibliography{references}

\newpage
\appendix
\onecolumn
\renewcommand{\thefigure}{\thesection\arabic{figure}}
\renewcommand{\thetable}{\thesection\arabic{table}}

\section{Incorporating Narrative and Cinematic Principles into \system}\label{app:books}
\setcounter{table}{0}
To inform our paradigm of generative expansion (\autoref{sec:design_goals}) and the design of \system (\autoref{sec:system_design}), we synthesized established narrative and cinematic principles from 9 filmmaking texts.
These insights were then translated into system prompts, which are used as additional context when generating new content and script suggestions (\autoref{app:pipeline}).
\tableBooks

Script suggestions are generated and categorized based on the following narrative categories:
\tableCategories

\newpage
\section{\system AI-Powered Features and Pipeline}\label{app:pipeline}
\setcounter{figure}{0}
\setcounter{table}{0}
In \autoref{tab:system_features}, we include an overview of all AI-powered features in our system.
\autoref{fig:suggestion_pipeline} illustrates the prompting pipeline used to generate script suggestions in more detail (similar to \autoref{fig:generative_pipeline} from \autoref{sec:implementation}).
Narrative and cinematic principles are derived from our book analysis (\autoref{app:books}), and included as prompts (see Supplementary Materials).

\tableFeatures

\figureSuggestionPipeline

\clearpage
\section{Additional Interface Screenshots}
\setcounter{figure}{0}
\autoref{fig:compare} shows what \system looks like when comparing story versions. 
This is not a major feature, but creators found it useful to visualize multiple versions at the same time and receive a personalized analysis of strengths and weaknesses 
(\autoref{sec:rq2}).

\figureCompare

\autoref{fig:new_scene} demonstrates how \system can intelligently suggest new transition scenes between existing ones, contextualized to the user's video story. 

\figureNewScene

\autoref{fig:controls} illustrates our image editor and video generation panels, which contain structured prompt fields to customize and refine shots.
In the video generation panel, creators also enjoyed using the spatial annotation feature to steer model generations.

\figureControls

\clearpage
\section{Additional Examples of \system in Action}
\setcounter{figure}{0}
Below, \autoref{fig:storyboard} and \autoref{fig:creative_storytelling} show examples of how \system helped our user study participants expand their video stories in creative, contextualized ways (\autoref{sec:results}).

\figureStoryboard

\figureCreativeStorytelling

\end{document}